# Gravity-induced wavefunction-collapse in a temporally expanding spacetime

*Garrelt Quandt-Wiese*
*Schlesierstr. 16*
*64297 Darmstadt*
*Germany*
*garrelt@quandt-wiese.de*
*http://www.quandt-wiese.de*

A gravity-induced approach to wavefunction collapse based on semiclassical gravity is enhanced by the hypothesis of a temporally expanding spacetime, which leads to a collapse model that can resolve the conflict between quantum nonlocality and relativity. It is postulated that the spacetime region on which the evolution of the state vector exists is bounded towards the future by a border that is dynamically moving towards the future, and at which the state vector must fulfil a boundary condition. Wavefunction collapse is represented in such a way that the evolution of the state vector changes abruptly at critical spacetime expansions to an evolution resembling a classical trajectory. This can explain the correlations in EPR experiments without coming into conflict with relativity, since the evolution of the state vector before and after the abrupt change is governed solely by local physical laws. This model leads to the same lifetimes of superpositions as the gravity-based approaches of Diósi and Penrose, and is characterised by the facts that energy is conserved at collapse and that the reduction point in time does not vary statistically. Some unique features of the model are that it naturally leads to stochastic behaviour and that it can predict reduction probabilities. It explains why all experiments performed so far behave in agreement with Born's rule, due to a property that they have in common. This gives rise to new experiments for checking Born's rule, which can be realised in the short term.

## 1. Introduction

Even after more than 100 years of quantum theory, the collapse of the wavefunction remains one of the unresolved problems of physics. Despite the many approaches to collapse that have been developed and discussed [1-4], the following two fundamental questions remain open: firstly, what is the physical origin of collapse, and secondly, how can we resolve the conflict between the nonlocal nature of quantum theory, as manifested by the observation of quantum correlations in EPR experiments, and the local nature of relativity, whose spacetime symmetries are the basis of all established physical laws?

Regarding the first question, the candidate most often discussed is gravity [4,5]. Diósi proposed a noise-based dynamical reduction model [6,7], whereas Penrose suggested a mechanism based on heuristic arguments [9,10]. Both approaches lead to the same orders of magnitude for the collapse time, which can be estimated with the so-called Diósi-Penrose criterion [19], and this is often used when discussing the feasibility of experimental proposals for measuring collapse [10-18].

The nonlocal nature of quantum theory was demonstrated by Aspect in 1982 [20] by measuring the violation of Bell's inequalities in EPR experiments. The consequence that this result leads to a fundamental conflict with relativity was ignored for a long time [21]. The Free Will theorem of Conway and Kochen in 2006, which showed that in special EPR experiments, with free choice of measurement, information about the choices must travel infinitely fast between the measurement partners [22-24], made the conflict obvious, and this was additionally reinforced by problems arising from the formulation of relativistic collapse models [29-32].

---

To overcome the conflict between quantum nonlocality and relativity, we propose a new view of the dynamical evolution of physical systems. We postulate that the spacetime region on which quantum fields exist and on which the evolution of the state vector can be regarded is bounded towards the future by a border that is dynamically moving towards the future. For every expansion of spacetime, a new solution for the evolution of the state vector in spacetime must be found that must fulfil a boundary condition at the border of spacetime. This defines a new causality chain, which is ordered by the expansion of spacetime.

The second pillar of our collapse model is the assumption of a classical, non-quantised spacetime known as semiclassical gravity [33,34]. Semiclassical gravity leads to a competition between superposed states for the curvature of spacetime, since as soon as superposed states have different mass distributions, they prefer differently curved spacetimes, according to general relativity, but must share a common spacetime geometry.

These two assumptions (a temporally expanding spacetime and semiclassical gravity) give results that go beyond the unitary evolution of quantum theory. As soon as the mass distributions of the superposed states are different, the frequencies of the wavefunctions of these states must slightly deviate from what is expected from the unitary evolution. With the help of these frequency deviations, a mechanism for the collapse can be derived, which leads to abrupt changes in the evolution of the wavefunction. With these abrupt changes, which are only possible at certain critical expansions of spacetime, the evolution of the wavefunction changes from an evolution that evolves into a superposition to an evolution that resembles a classical trajectory. This result for the collapse can explain the quantum correlations in EPR experiments without provoking a conflict with relativity, since the evolution of the wavefunction before and after the abrupt change is governed solely by local physical laws.

Some unique features of our collapse model are that it naturally leads to stochastic behaviour, and that it can predict reduction probabilities. The smallest fluctuations, which have their origin in the permanent expansion of spacetime, determine which classical trajectory the evolution of the wavefunction will take after the abrupt change at the critical expansions of spacetime. Our model can explain why all of the experiments performed so far behave in agreement with Born's rule, based on a property that they have in common. This property is that we can never locally distinguish more than two states, for example corresponding to “detection” or “no detection” of a particle at a given location. This finding gives rise to new experiments generating more than two locally distinguishable states. We propose to transfer a solid into a superposition of three states, with the solid in a slightly different position in each of the states. For this experiment, our collapse model predicts deviations from Born’s rule, which has the far-reaching consequence of the option to enhance EPR experiments with faster-than-light signalling. However, in the same way as for our explanation of quantum correlations, this prediction does not provoke a conflict with relativity in the physical context assumed here.

---

This paper is structured as follows. In Sections 2 to 6, our model is first developed for superpositions of two states. In Sections 2 and 3, we introduce the hypothesis of a temporally expanding spacetime and semiclassical gravity, in order to derive the collapse mechanism in Section 4 on the basis of these results. In Section 5, we extend our model to predict reduction probabilities and derive Born’s rule. In Section 6, we transfer the derived results into a covariant form, and show that our model can be considered to be relativistic. In Section 7, we then extend our model for superpositions for more than two states in order to apply it to typical quantum mechanical experiments in Section 8 and show why they always behave in accordance with Born's rule. Finally, in Section 9, we discuss an experiment, in which a solid is transferred into a three-state superposition, and for which our model predicts deviations from Born's rule, and show how this result can be used for faster-than-light signalling. The discussion in Section 10 ends with an overview of possibilities for experimental verification of the proposed *Temporally Expanding Spacetime approach* to wavefunction collapse.

## 2. Temporally expanding spacetime

In this section, we introduce the postulate of a temporally expanding spacetime, which leads to a new procedure for calculating the evolution of physical systems.

### 2.1. Postulate of a temporally expanding spacetime

Figure 1 illustrates the idea of a temporally expanding spacetime. It is assumed that the spacetime region on which quantum fields exist and on which the wavefunction’s evolution can be considered is bounded towards the future by a spacelike hypersurface $\bar{\sigma}$, referred to here as the *spacetime border*. The spacetime border is dynamically moving towards the future, and this evolution is parameterised by the so-called *spacetime expansion parameter* $\bar{\tau}$. The spacetime expansion parameter itself is not a physical observable quantity, and can be chosen to be dimensionless. It is assumed that the evolution of the spacetime border $\bar{\sigma}(\bar{\tau})$ is a sequence of time-ordered spacelike hypersurfaces, which means that no point of $\bar{\sigma}(\bar{\tau}_2)$ is within the past light-cone of a point on a preceding hypersurface $\bar{\sigma}(\bar{\tau}_1)$.

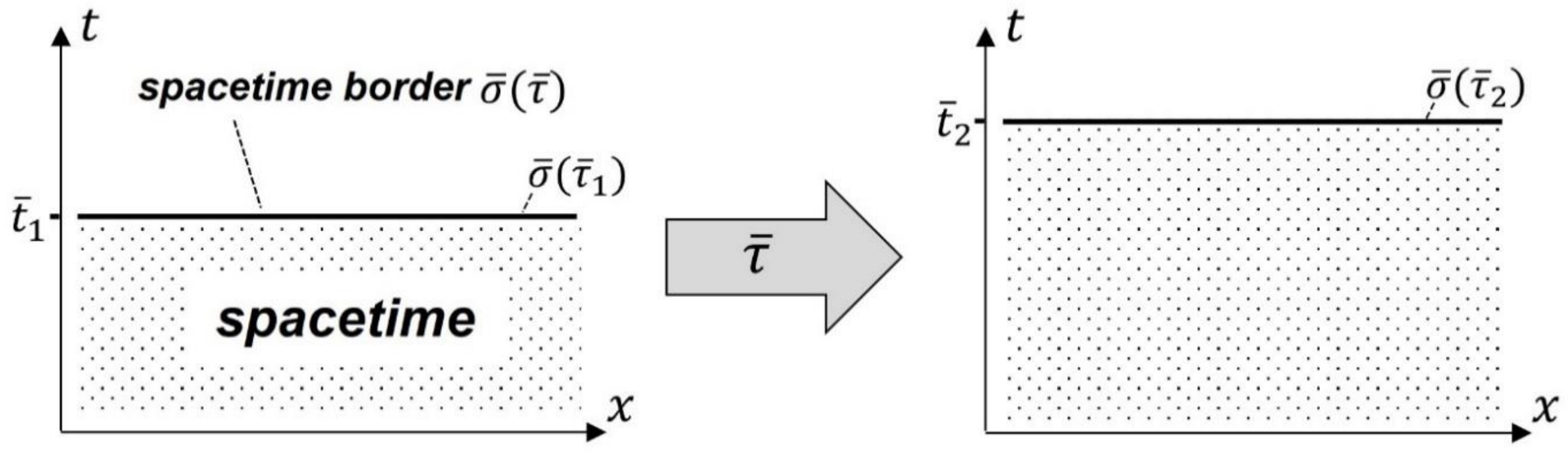

**Figure 1**. Illustration of a temporally expanding spacetime. The spacetime in which quantum fields exist and on which the evolution of the wavefunction can be considered ends at a *spacetime border* $\bar{\sigma}$, which is dynamically moving as function of the *spacetime expansion parameter* $\bar{\tau}$ towards the future, as shown to the right for a larger value of $\bar{\tau}$.

The discussion of most aspects of our collapse model can be simplified by assuming that the spacetime border is a plane hypersurface for which the propagation is aligned with the rest frame of the experiment. In this case, spacetime ends at a time instant $\bar{t}$ in this frame, as shown in Figure 1, and the spacetime expansion parameter $\bar{\tau}$ can be expressed based on this point in time $\bar{t}$ (i.e. $\bar{\tau} \rightarrow \bar{t}$).
With the spacetime border, we explicitly introduce into our model a distinction between the past and the future, where the past corresponds to the spacetime region before the spacetime border ($t < \bar{t}$), and the future will take place in regions of spacetime that the spacetime border has not yet reached.

### 2.2. Procedure for calculating the evolution of the wavefunction and a new definition of the causality chain

In the Temporally Expanding Spacetime approach, we assume that the state vector $|\psi >$ must fulfil a boundary condition at the spacetime border in such a way that the phase of $|\psi >$ is pinned to a fixed value at this border. This leads to a new procedure for calculating the evolution of the wavefunction. For each position of the spacetime border $\bar{\sigma}(\bar{t})$, a solution must be found for the evolution of the state vector $|\psi(t) >_{\bar{t}}$ on the spacetime region before $\bar{t}$, which fulfils the boundary condition at the spacetime border, i.e. for $t = \bar{t}$. This leads to a new definition of the causality chain in physics. While in relativistic physics, the state of a system can be followed up on any sequence of time-ordered spacelike hypersurfaces, the causality chain is now ordered by the expansion of spacetime, i.e. by the spacetime expansion parameter $\bar{t}$. When the spacetime border moves from $\bar{t}$ to $\bar{t} + d\bar{t}$, we calculate a new solution $|\psi(t) >_{\bar{t}+d\bar{t}}$ based on the earlier solution $|\psi(t) >_{\bar{t}}$, which now satisfies the boundary condition for $t = \bar{t} + d\bar{t}$. This defines a causality chain, which is ordered by the spacetime expansion parameter $\bar{t}$.

Without gravity, i.e. for a flat spacetime, this procedure for calculating the evolution of the wavefunction leads to the same results as the standard quantum theory. The evolution of the wavefunction in spacetime up to the spacetime border can then be calculated with the unitary evolution. With the introduction of semiclassical gravity, however, new behaviours emerge. At certain critical expansions of spacetime, the evolution of the wavefunction abruptly changes from an evolution that evolves into a superposition to an evolution that resembles a classical trajectory. This explanation for collapse leads to a new way of looking at the collapse of the wavefunction. While one normally assumes that the wavefunction changes at a point in time during collapse, i.e. on the space-like hypersurface that can be assigned to this point in time, the wavefunction now changes over an entire spacetime region. The collapse of the wavefunction thus becomes an event in space and time.

# 3. Semiclassical gravity and frequency detuning of paths

In this section, we first introduce semiclassical gravity and then show that if the state vector evolves into a superposition of states with different mass distributions and it must fulfil a boundary condition at the spacetime border, then the frequencies of the states' wavefunctions are slightly detuned with respect to the unitary evolution. This result will form the basis for our derivation of the collapse mechanism in Section 4.

## 3.1. Experimental generation of a superposition of states with different mass distributions

A superposition of two states with different mass distributions can be generated using the experiment shown on the left of Figure 2. In this experiment, a single photon is split by a beam splitter, and is measured by the detector on the right. When this detector measures the photon, it shifts a rigid body to the right, as shown in the figure. This leads to a superposition of two states with mass distributions $\rho_1(\boldsymbol{x})$ and $\rho_2(\boldsymbol{x})$, as shown in Figure 2, for which the evolution of the state vector can be written as:

$$|\psi(t)>= c_1|\psi_1(t)> + c_2|\psi_2(t)>, \tag{1}$$

with $|c_1|^2 + |c_2|^2 = 1$. The right-hand side of the figure shows the evolution of the state vector in configuration space, where in our discussion the state vector always describes the complete system, consisting here of the photon, the beam splitter, the detector, and the rigid body. At time $t_s$, when the photon enters the beam splitter, the evolution of the state vector splits into two well-separated wave packets $|\psi_1(t)>$ and $|\psi_2(t)>$, denoted in Figure 2 as paths 1 and 2.

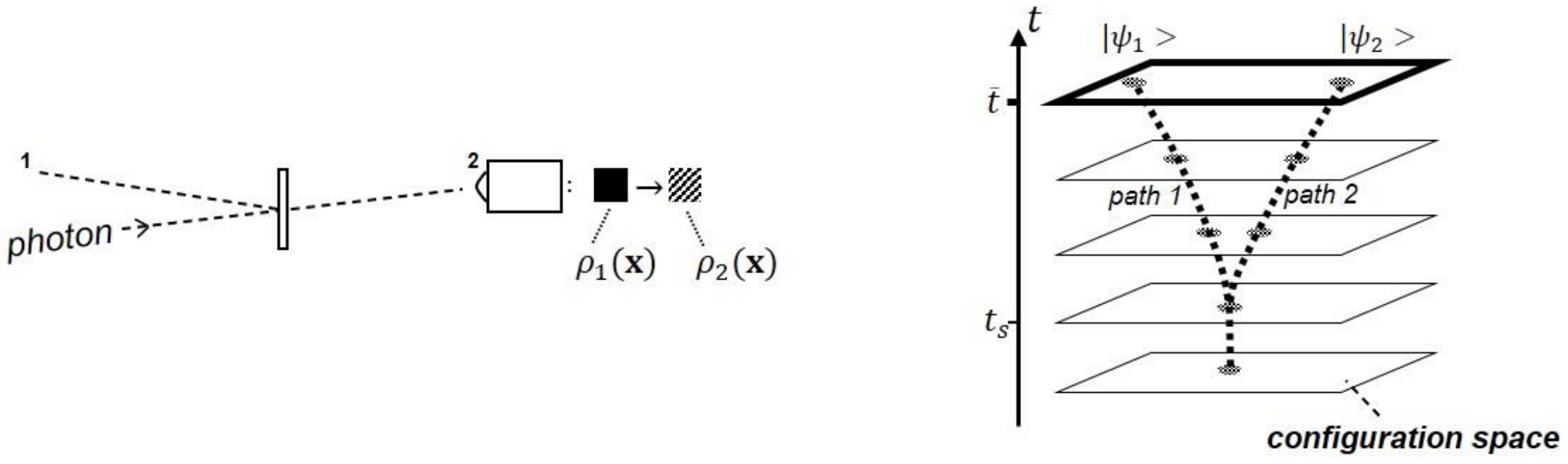

**Figure 2.** *Left:* Experimental generation of a superposition of two states with different mass distributions $\rho_1(\boldsymbol{x})$ and $\rho_2(\boldsymbol{x})$. When the detector measures the photon, it shifts the rigid body to the right. *Right:* Illustration of the evolution of the state vector in configuration space, which splits into two well-separated wave packets at $t_s$, when the photon enters the beam splitter.

### 3.2. Semiclassical gravity

The question of whether or not the gravitational field must be quantised is the subject of open scientific debate [33-38]. We assume a classical, non-quantised spacetime, which is known as semiclassical gravity [33,34]. Semiclassical gravity provides a starting point for a collapse mechanism, since it leads to a competition between superposed states for the curvature of spacetime. As soon as superposed states have different mass distributions, they prefer differently curved spacetimes, according to general relativity, but must share a common, classical spacetime geometry. However, semiclassical gravity alone is not sufficient for an explanation of wavefunction collapse, which is known from studies of the Newton-Schrödinger equation [39,40] describing semiclassical gravity in the Newtonian limit [33].

In the Newtonian limit of general relativity, the calculation of the metric field $g_{\mu\nu}(x)$ from the energy momentum tensor field $T_{\mu\nu}(x)$ and Einstein's field equations can be restricted to the $g_{00}$-component, whose square root describes the derivation of physical time $s$ with respect to the time coordinate $x^0$ as [41]:

$$\frac{ds}{dx^0} = \sqrt{g_{00}}\,. \tag{2}$$

The square root of the $g_{00}$-component is then directly related to the gravitational potential $\Phi(\mathbf{x})$ by [41]:

$$\sqrt{g_{00}} \approx 1 + \frac{\Phi(\mathbf{x})}{c^2}\,, \tag{3}$$

which can be calculated from the mass distribution $\rho(\mathbf{x})$ by:

$$\Phi(\mathbf{x}) = -G \int d^3\mathbf{y}\, \frac{\rho(\mathbf{y})}{|\mathbf{x}-\mathbf{y}|}\,. \tag{4}$$

The mass distribution $\rho(\mathbf{x})$ in our experiment (Figure 2) can be determined from the wavefunction in (1) with the mass density operator $\hat{\rho}(\mathbf{x})$, which yields:

$$\rho(\mathbf{x}) = |c_1|^2 \rho_1(\mathbf{x}) + |c_2|^2 \rho_2(\mathbf{x})\,, \tag{5}$$

with

$$\rho_i(\mathbf{x}) = <\psi_i|\hat{\rho}(\mathbf{x})|\psi_i>. \tag{6}$$

The sharing of a common spacetime geometry under semiclassical gravity is, in the Newtonian limit, synonymous with the sharing of a common gravitational potential $\Phi_{com}(\mathbf{x})$. The common gravitational potential can be calculated using (4) and the mass distribution of the superposition in (5), which yields:

$$\Phi_{com}(\mathbf{x}) = |c_1|^2 \Phi_1(\mathbf{x}) + |c_2|^2 \Phi_2(\mathbf{x})\,, \tag{7}$$

where $\Phi_i(\mathbf{x})$ is the gravitational potential belonging to the mass distribution $\rho_i(\mathbf{x})$ of state $i$:

$$\Phi_i(\mathbf{x}) = -G \int d^3\mathbf{y} \frac{\rho_i(\mathbf{y})}{|\mathbf{x}-\mathbf{y}|} . \quad (8)$$

### 3.3. Gravitational energy of states in superpositions and the Diósi-Penrose criterion

In this section, we lay the groundwork for calculating the frequency detuning of the paths in Section 3.4 by a calculation of the gravitational energies of states in superpositions, which will lead us to the characteristic gravitational energy of the Diósi-Penrose criterion.

The gravitational energy $E_{G\,i}$ of a state $i$ in in the common gravitational potential $\Phi_{com}(\mathbf{x})$ of a superposition of $N$ states is given by:

$$E_{G\,i} = \int d^3\mathbf{x}\,\rho_i(\mathbf{x})\Phi_{com}(\mathbf{x}) = \int d^3\mathbf{x}\,\rho_i(\mathbf{x}) \sum_{j=1}^{N} |c_j|^2 \Phi_j(\mathbf{x}) . \quad (9)$$

From norm conservation ( $\sum_{i=1}^{N} |c_i|^2 = 1$), (9) can be written as:

$$E_{G\,i} = \int d^3\mathbf{x}\,\rho_i \Phi_i + \sum_{j\neq i} |c_j|^2 \int d^3\mathbf{x}\,\rho_i(\Phi_j - \Phi_i) . \quad (10)$$

Using the relation $\int d^3\mathbf{x}\,\rho_i\Phi_j = \int d^3\mathbf{x}\,\rho_j\Phi_i$, the terms in the sum in (10) can be transformed as:

$$\int d^3\mathbf{x}\,\rho_i(\Phi_j - \Phi_i) = -\frac{1}{2}\int d^3\mathbf{x}\,(\rho_j - \rho_i)(\Phi_j - \Phi_i) + \frac{1}{2}\int d^3\mathbf{x}\,(\rho_j\Phi_j - \rho_i\Phi_i) . \quad (11)$$

Based on this, we get the following result for $E_{G\,i}$:

$$E_{G\,i} = \int d^3\boldsymbol{x}\,\rho_i(\boldsymbol{x})\phi_i(\boldsymbol{x}) + \sum_{j\neq i} |c_j|^2 E_{DP\,ij} + \sum_{j\neq i} |c_j|^2 \left(E_{G\,self\,j} - E_{G\,self\,i}\right) , \quad (12)$$

where $E_{DP\,ij}$ is defined as:

$$E_{DP\,ij} = -\frac{1}{2}\int d^3\mathbf{x}\,(\rho_j(\mathbf{x}) - \rho_i(\mathbf{x}))(\Phi_j(\mathbf{x}) - \Phi_i(\mathbf{x})) , \quad (13)$$

and $E_{G\,self\,i}$ as:

$$E_{G\,self\,i} = \frac{1}{2}\int d^3\mathbf{x}\,\rho_i(\mathbf{x})\Phi_i(\mathbf{x}) . \quad (14)$$

$E_{G\,self\,i}$ is the gravitational self-energy of state $i$, i.e. the gravitational energy resulting

from its mass distribution $\rho_i(\mathbf{x})$, where the factor $\frac{1}{2}$ in (14) ensures that the gravitational energy between two masses ($E_G = -Gm_1m_2/r$) is not calculated twice during the integration.

$E_{DP\,ij}$ is denoted here as the *Diósi-Penrose energy*, which can be transformed via (8) into the better-known form:

$$E_{DP\,12} = \frac{G}{2}\int d^3\boldsymbol{x}d^3\boldsymbol{y}\frac{(\rho_2(\boldsymbol{x}) - \rho_1(\boldsymbol{x}))(\rho_2(\boldsymbol{y}) - \rho_1(\boldsymbol{y}))}{|\boldsymbol{x} - \boldsymbol{y}|}\,. \tag{15}$$

The Diósi-Penrose energy plays a central role in the gravity-based approaches to wavefunction collapse put forward by Diósi [6,7] and Penrose [9,10]. With this energy, the lifetime of a two-state superposition can be estimated by the following rule of thumb:

$$T_G \approx \frac{\hbar}{E_{DP\,12}}, \tag{16}$$

which is sometimes referred to as the Diósi-Penrose criterion [19]. The Diósi-Penrose criterion is often used to assess experimental proposals for measuring wavefunction collapse [10-18].

It is important to note that the Diósi-Penrose energy is always positive, which follows directly from Penrose's approach [10]. From (13), it follows that the Diósi-Penrose energy is minus the gravitational self-energy of the mass difference of the superposed states $(\rho_2(\boldsymbol{x}) - \rho_1(\boldsymbol{x}))$. In the context of semiclassical gravity, a further illustration of the Diósi-Penrose energy can be given by calculating the gravitational self-energy of the superposition ($E_{G\,self} = \frac{1}{2}\int d^3\mathbf{x}\rho(\mathbf{x})\Phi(\mathbf{x})$), which yields[1]:

$$E_{G\,self} = |c_1|^2 E_{G\,self\,1} + |c_2|^2 E_{G\,self\,2} + |c_1|^2|c_2|^2 E_{DP\,12}\,. \tag{17}$$

Here, the Diósi-Penrose energy describes the increase in the gravitational self-energy of the superposition with respect to $|c_1|^2 E_{G\,self\,1} + |c_2|^2 E_{G\,self\,2}$, which is the value expected when every state resides in its own gravitational potential resulting from its mass distribution. The increase in the gravitational self-energy by $|c_1|^2|c_2|^2 E_{DP\,12}$ follows from the fact that the states must share a common gravitational potential under semiclassical gravity.

### 3.4. Frequency detuning of paths

In this section, we calculate the frequency detuning of the paths resulting from the two postulates of our model, a temporally expanding spacetime and semiclassical gravity. The frequency detunings are slight deviations in the frequencies of the wavefunctions of the paths with respect to the unitary evolution, which are of central importance for the derivation of the collapse mechanism in Section 4. The frequency detuning of the paths has its physical cause in the fact that the wavefunctions of the two paths must, on the one hand, fulfil the boundary condition at the spacetime border and, on the other hand, be in phase at the splitting point of the superposition. Both conditions together can only be fulfilled if frequency detunings are introduced for the paths.

[1] This result follows from inserting the equations in (5) and (7) into $E_{G\,self} = \frac{1}{2}\int d^3\mathbf{x}\rho\Phi$ and a short calculation similar to the one in the footnote to (59), or using $E_{G\,self} = \frac{1}{2}\sum_i |c_i|^2 E_{G\,i}$. For a superposition of $N$ states, we get: $E_{G\,self} = \sum_i |c_i|^2 E_{G\,self\,i} + \sum_{ij}\frac{1}{2}|c_i|^2|c_j|^2 E_{DP\,ij}$.

For the calculation of the frequency detuning of the paths, we investigate how the common spacetime geometry of the two paths changes when intensity is shifted between them. A change in the spacetime geometry causes the number of oscillations that the wavefunctions of the two paths have between the splitting point of the superposition and the spacetime border to change slightly, because the available spacetime volume changes. Since the phases of the paths' wavefunctions are pinned to a fixed value at the spacetime border (cf. Section 1), this can lead to phase shifts between the wavefunctions of the two paths at the splitting point of the superposition, which can be resolved by introducing a frequency detuning for the paths. To calculate the absolute frequency detuning of a path, we first set its intensity to one, so that the intensity of the other path is zero, and the system is not in any superposition, and thus no frequency detuning can occur. Then we shift intensity from the considered path in favour of the other path until the considered path has the desired intensity. From this intensity shift, we calculate the absolute frequency detuning of the considered path.

We first consider this effect for the wavefunction of a single atom. For an intensity shift of $dI$ from path 1 in favour of path 2 ($|c_1|^2 \to |c_1|^2 - dI$, $|c_2|^2 \to |c_2|^2 + dI$), the common gravitational potential of the superposition changes as $\Delta\Phi_{com}(\mathbf{x}) = dI(\Phi_2(\mathbf{x}) - \Phi_1(\mathbf{x})$ (cf. (7)). According to (3) and (2), an increase of $\Delta\Phi_{com}(\mathbf{x})$ increases the physical time $\Delta s$ corresponding to the interval of the time coordinate $\Delta x^0$ by ($\Delta s \to \Delta s + \delta s$):

$$\delta s = \frac{\Delta\Phi_{com}(\mathbf{x})}{c^2}\Delta x^0 \,. \tag{18}$$

Due to this effect, the phase angle for the wavefunction of a single atom with frequency $\hbar\omega_{atom} = m_{atom}c^2$ at location $\mathbf{x}_{atom}$ rotates in the interval $\Delta x^0$ through an amount $\delta\varphi$ further than before the potential change, which can be expressed as:

$$\hbar\delta\varphi = m_{atom}\Delta\Phi_{com}(\mathbf{x}_{atom})\Delta x^0 \,. \tag{19}$$

The total phase change $\delta\varphi_1$ for the wavefunction of path 1 due to the potential change $\Delta\Phi_{com}(\mathbf{x})$ follows from this result when we take all of the atoms in our experiment into account, which yields: $\hbar\delta\varphi_1 = \int d^3\mathbf{x}\rho_1(\mathbf{x})\Delta\Phi_{com}(\mathbf{x}) \cdot \Delta x^0$. When the phase of the wavefunction is pinned to a fixed value at the spacetime border $\bar{t}$, as assumed in our model, all of the phase changes $\delta\varphi_1$ throughout the time interval $[t_s, \bar{t}]$ lead to a phase change $\Delta\varphi_1 = -\int_{t_s}^{\bar{t}} dx^0 \delta\varphi_1(x^0)$ at the splitting point of the superposition $t_s$. This leads to the following phase changes for the wavefunctions of paths 1 and 2 at $t_s$ due to the change in potential $\Delta\Phi_{com}(\mathbf{x})$:

$$\hbar\Delta\varphi_1 = -\int_{t_s}^{\bar{t}} dx^0 \int d^3\mathbf{x}\rho_1(\mathbf{x}, x^0)\Delta\Phi_{com}(\mathbf{x}, x^0)$$
$$\hbar\Delta\varphi_2 = -\int_{t_s}^{\bar{t}} dx^0 \int d^3\mathbf{x}\rho_2(\mathbf{x}, x^0)\Delta\Phi_{com}(\mathbf{x}, x^0) \quad . \tag{20}$$

The terms $\int d^3\mathbf{x}\rho_i(\mathbf{x})\Delta\Phi_{com}(\mathbf{x})$ in these equations follow from (9) and (12), and become with $d|c_1|^2 = -dI$ and $d|c_2|^2 = dI$:

$$\int d^3\mathbf{x}\rho_1(\mathbf{x})\Delta\Phi_{com}(\mathbf{x}) = dI\,(+E_{DP\,12} + E_{\mathrm{G}\,self\,2} - E_{\mathrm{G}\,self\,1})$$
$$\int d^3\mathbf{x}\rho_2(\mathbf{x})\Delta\Phi_{com}(\mathbf{x}) = dI\,(-E_{DP\,12} + E_{\mathrm{G}\,self\,2} - E_{\mathrm{G}\,self\,1}) \quad . \tag{21}$$

To calculate the absolute frequency detuning of a path 1, we change its intensity from one to $|c_1|^2$, as described at the beginning, which corresponds to an intensity shift from path 1 in favour of path 2 of $dI = |c_2|^2$ ($|c_1|^2 + |c_2|^2 = 1$). Inserting (21) with $dI = |c_2|^2$ into (20), we get the following phase changes for paths 1 and 2 at $t_s$:

$$\hbar\Delta\varphi_1 = -|c_2|^2 \int_{t_s}^{\bar{t}} dt\,(+E_{DP\,12}(t) + E_{G\,self\,2}(t) - E_{G\,self\,1}(t))$$
$$\hbar\Delta\varphi_2 = -|c_2|^2 \int_{t_s}^{\bar{t}} dt\,(-E_{DP\,12}(t) + E_{G\,self\,2}(t) - E_{G\,self\,1}(t)) \quad . \tag{22}$$

For different gravitational self-energies of the states ($E_{G\,self\,2} \neq E_{G\,self\,1}$) and a vanishing Diósi-Penrose energy between the states ($E_{DP\,12} = 0$), the phases of the wavefunctions of both paths change in the same way ($\Delta\varphi_1 = \Delta\varphi_2$), and stay in phase. Hence, we need no frequency detuning with respect to the unitary evolution to keep the phases of the wavefunctions of paths 1 and 2 in phase. However, for a non-vanishing Diósi-Penrose energy ($E_{DP\,12} \neq 0$) and identical gravitational self-energies ($E_{G\,self\,2} = E_{G\,self\,1}$), the phases of paths 1 and 2 change in opposite directions ($\Delta\varphi_1 = -\Delta\varphi_2$). To keep the phases of the wavefunctions of paths 1 and 2 at $t_s$ in phase ($\Delta\varphi_1 = \Delta\varphi_2$), the frequency of path 1 must be detuned. The additional phase change that must result from this detuning $\Delta\varphi_{det\,1}$ is given by:

$$\Delta\varphi_{det\,1} = \frac{\Delta\varphi_2 - \Delta\varphi_1}{2} = \frac{|c_2|^2}{\hbar}\int_{t_s}^{\bar{t}} dt\, E_{DP\,12}(t)\,, \tag{23}$$

which means that the frequency of path 1 must be changed over $[t_s, \bar{t}]$ on average by:

$$\overline{\Delta\omega_1} = -\frac{\Delta\varphi_{det\,1}}{\bar{t} - t_s}\,. \tag{24}$$

The frequency detunings of the paths are expressed here in the form of so-called *detuning actions*. The detuning action of path 1 is defined by:

$$S_{det\,1}(\bar{t}) \equiv -\hbar\overline{\Delta\omega_1}(\bar{t}-t_s) \;=\; |c_2|^2 \int_{t_s}^{\bar{t}} dt\, E_{DP\,12}(t)\,. \tag{25}$$

The detuning actions describe the action changes resulting from the frequency detuning of the paths. We define the detuning actions in such a way that they are positive (note that in (25), $E_{DP\,12}$ is positive), since only the amount of detuning will play a role in the derivation of the collapse mechanism. The frequency detunings $\overline{\Delta\omega_i}$ corresponding to the detuning actions $S_{det\,i}$ are negative (cf. (24) and (23)).

The frequency detuning and the detuning action of path 2, $\overline{\Delta\omega_2}$ and $S_{det\,2}$, can be calculated with the same procedure as for path 1, by setting its intensity first to one and then decreasing its intensity from one to $|c_2|^2$, which leads to:

$$S_{det\,2}(\bar{t}) \equiv -\hbar\overline{\Delta\omega_2}(\bar{t}-t_s) \;=\; |c_1|^2 \int_{t_s}^{\bar{t}} dt\, E_{DP\,12}(t)\,. \tag{26}$$

### 3.5. Competition action

The detuning actions of paths 1 and 2 can be written as:

$$\begin{aligned} S_{det\,1}(\bar{t}) &= |c_2|^2\, S_{comp\,12}(\bar{t}) \\ S_{det\,2}(\bar{t}) &= |c_1|^2\, S_{comp\,12}(\bar{t})\,, \end{aligned} \tag{27}$$

where the action $S_{comp\,12}(\bar{t})$ is given by:

$$S_{comp\,12}(\bar{t}) \equiv \int_{t_s}^{\bar{t}} dt\, E_{DP\,12}(t)\,. \tag{28}$$

The action $S_{comp\,12}(\bar{t})$ is referred to as the *competition action* between the paths. According to (27), it determines the strength of the detunings of the paths, and can be regarded as a measure of how strongly the paths compete for spacetime geometry.

An important characteristic of result (27) for the paths' detuning actions is that the detuning action for one path is proportional to the intensity of the other path with which it is competing for spacetime geometry ($S_{det\,1} \sim |c_2|^2$, $S_{det\,2} \sim |c_1|^2$). This result follows intuitively from the argument that the detuning of a path depends on how much its preferred spacetime geometry deviates from the spacetime geometry of the superposition. The stronger the intensity of the other, competing path, the more the spacetime geometry of the superposition is determined by this path, and the more the preferred spacetime geometry of the first path deviates from the spacetime geometry of the superposition.

Our definition of the competition action in (28) can also be generalised when defining the mass distributions of paths 1 and 2 $\rho_1(\boldsymbol{x})$ and $\rho_2(\boldsymbol{x})$ for $t < t_s$. This can be done by defining their mass distributions for $t < t_s$ based on the mass distribution of their common root wave-packet, which splits at $t_s$ into paths 1 and 2 (cf. Figure 2). The integration in (28) can then start at any point in time before $t_s$, which leads to:

$$S_{comp\ 12}(\bar{t}) \equiv \int_{..}^{\bar{t}} dt\ E_{DP\ 12}(t)\ . \tag{29}$$

This generalised definition of the competition action leads to the same results as (28), since the Diósi-Penrose energy between identical states vanishes (cf. the equation in (13)). By inserting the Diósi-Penrose energy (13) into (29), the competition action can be written as an integral over spacetime up to the spacetime border $\bar{\sigma}(\bar{t})$ as:

$$S_{comp\ 12}(\bar{t}) = -\frac{1}{2}\int_{..}^{\bar{\sigma}(\bar{t})} \frac{d^4x}{c}\left(\rho_2(x) - \rho_2(x)\right)\left(\Phi_2(x) - \Phi_1(x)\right)\ . \tag{30}$$

# 4. Wavefunction collapse and quantum correlations

In this section, we derive the collapse mechanism of the Temporally Expanding Spacetime approach using the frequency detuning of the paths from Section 3, and discuss the quantum correlations in EPR experiments.

## 4.1. Stability analysis of the evolution of the wavefunction

In this section, we carry out a stability analysis of the evolution of the wavefunction based on the frequency detuning of the paths. For convenience, we abbreviate the intensities of the paths $|c_i|^2$ in the following as:

$$I_i \equiv |c_i|^2\ . \tag{31}$$

In our stability analysis, we consider the intensity fluctuations for the two paths in Figure 2, $dI_1$ and $dI_2$ with $dI_1 + dI_2 = 0$ (norm conservation), to which the system reacts according to the result in (27) with the following changes in the detuning actions for the paths:

$$\begin{aligned} dS_{det\ 1} &= S_{comp\ 12}(\bar{t})\ dI_2 \\ dS_{det\ 2} &= S_{comp\ 12}(\bar{t})\ dI_1 \end{aligned}\ . \tag{32}$$

According to the discussion in Section 3.4, the evolution of the wavefunction reacts to these changes in the detuning actions with frequency changes such as $\psi \sim e^{-i\omega t} \rightarrow \psi \sim e^{-i(\omega+\Delta\omega)t}$, with $\hbar\Delta\omega \approx -dS_{det}/(\bar{t} - t_s)$. Another reaction to a change in the detuning of a path could be an exponential increase or decrease, such as $\psi \sim e^{-i\omega t} \rightarrow \psi \sim e^{-i\omega t \pm t/\tau}$, where $\hbar/\tau \approx dS_{det}/(\bar{t} - t_s)$, although this is not expected since it violates norm conservation. However, from this consideration, we conclude that a path tries to decrease/increase its intensity when its detuning action is increased/decreased. Since due to norm conservation, the intensity of a path must remain constant over the time interval $[t_s, \bar{t}]$, the only option is that the intensity is shifted in favour of the other path at the splitting point of the superposition $t_s$. The amount of such an intensity shift $dI$ can be estimated using $I = |\psi|^2$, $I(t) = Ie^{-2(t-t_s)/\tau}$, $\hbar/\tau = dS_{det}/(\bar{t} - t_s)$ and $I - dI = \int_{t_s}^{\bar{t}} dtI(t)$ as:

$$dI = I \, \frac{dS_{det}}{\hbar}, \tag{33}$$

where this result depends only on the change in the path's detuning action $dS_{det}$, and not on its length $(\bar{t} - t_s)$. The intensity shifts from path 1 in favour of path 2 $dI_{1\to 2}$ and vice versa ($dI_{2\to 1}$) resulting from changes $dS_{det\,1}$ and $dS_{det\,2}$ in their detuning actions follow from this result and (32):

$$\begin{aligned} dI_{1\to 2} &= I_1 \, \frac{dS_{det\,1}}{\hbar} = I_1 \, \frac{S_{comp\,12}(\bar{t})}{\hbar} dI_2 \\ dI_{2\to 1} &= I_2 \, \frac{dS_{det\,2}}{\hbar} = I_2 \, \frac{S_{comp\,12}(\bar{t})}{\hbar} dI_1 \end{aligned} . \tag{34}$$

For $dI_1 = -\,dI_{1\to 2} + dI_{2\to 1}$ and $dI_2 = dI_{1\to 2} - dI_{2\to 1}$, we get the following conditional equation for the possible intensity changes $dI_1$ and $dI_2$ in paths 1 and 2:

$$\begin{aligned} dI_1 &= \frac{S_{comp\,12}(\bar{t})}{\hbar} (I_2 \, dI_1 - I_1 dI_2) \\ dI_2 &= \frac{S_{comp\,12}(\bar{t})}{\hbar} (I_1 \, dI_2 - I_2 \, dI_1) \end{aligned} . \tag{35}$$

This conditional equation can be transformed with $dI_2 = -dI_1$ and $I_1 + I_2 = 1$ to:

$$\begin{pmatrix} dI_1 \\ dI_2 \end{pmatrix} = \frac{S_{comp\,12}(\bar{t})}{\hbar} \begin{pmatrix} 1 & 0 \\ 0 & 1 \end{pmatrix} \begin{pmatrix} dI_1 \\ dI_2 \end{pmatrix} . \tag{36}$$

This transformed conditional equation also allows for non-norm-conserving solutions ($dI_2 + dI_1 \neq 0$), but these are not solutions of our original conditional equation (35).

### 4.2. Collapse of the evolution of the wavefunction

From the transformed conditional equation (36), it follows that intensity changes in the paths are only possible when the competition action between the paths is exactly $\hbar$:

$$S_{comp\,12}(\bar{t}_C) = \hbar \, . \tag{37}$$

At this critical expansion of spacetime $\bar{t}_C$, the evolution of the wavefunction is expected to behave as follows. If there is a small intensity fluctuation in favour of path 2, for example, this fluctuation increases the detuning action of path 1 according to (32), which induces an intensity shift from path 1 in favour of path 2. This reinforces the initial intensity fluctuation in favour of path 2, which further increases the detuning action of path 1. Thus, we enter a self-reinforcing loop, which does not stop until the intensity of path 1 has completely vanished. In the same way, an intensity fluctuation in favour of path 1 leads to a complete vanishing of path 2. Thus, at the critical expansion of spacetime $\bar{t}_C$, we get abrupt changes in the wavefunction's evolution either in favour of path 1 or 2, as shown in Figure 3. In these evolutions, all particles of the system move approximately along classical trajectories. These evolutions of the state vector, in which all particles are evolving along classical trajectories, are referred to here as *classical scenarios*. For the classic scenarios shown in Figure 3, the photon is either completely reflected by or passes the beam splitter, as indicated in the insets of the figure representing the photon trajectories and the final position of the rigid body after detection. These abrupt changes in the evolution of the state vector from superpositions to classical scenarios describe the collapse of the wavefunction.

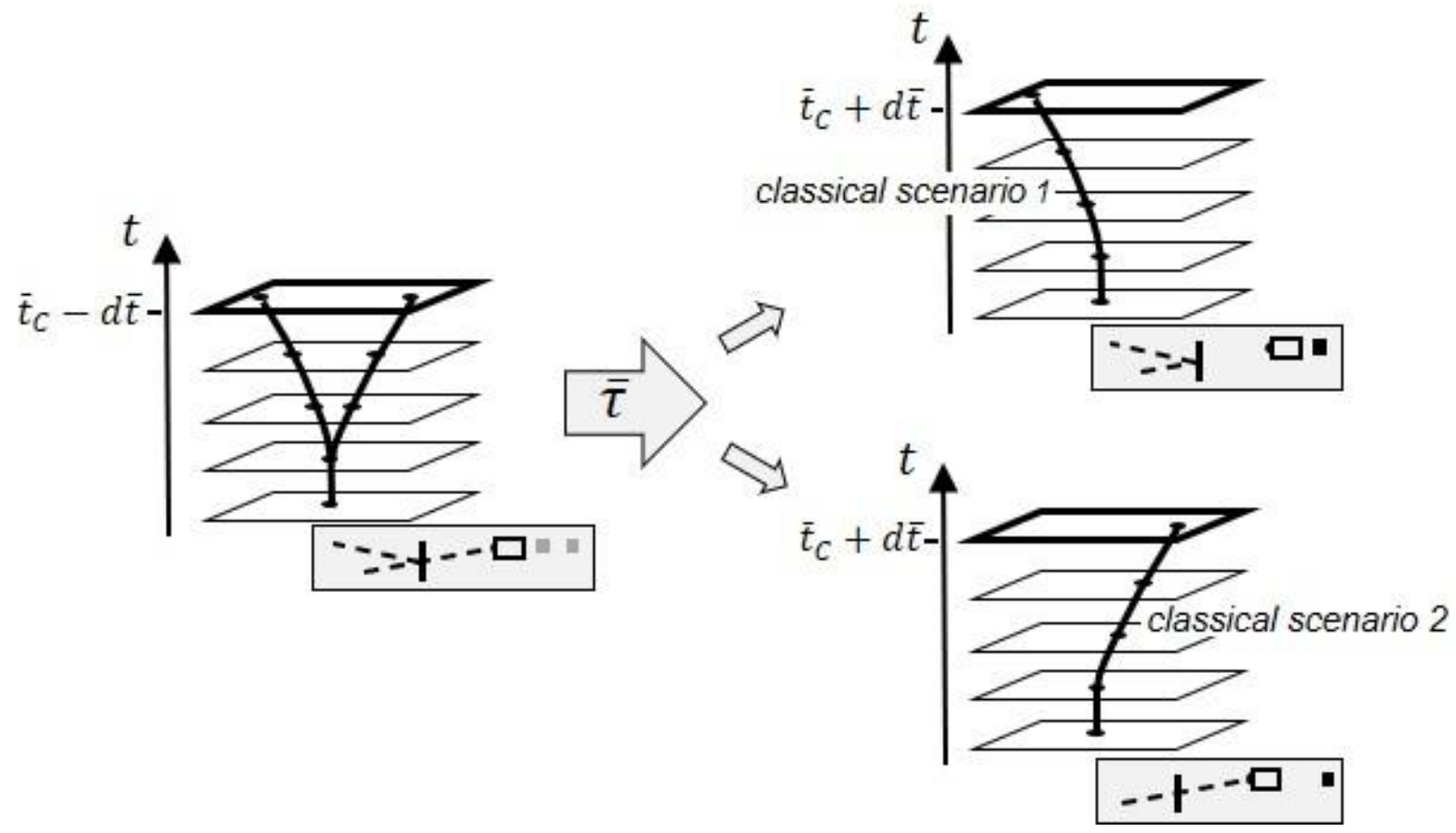

**Figure 3.** Collapse of the evolution of the wavefunction when the spacetime border passes the critical expansion point in spacetime $\bar{t}_C$. The evolution of the wavefunction in the experiment in Figure 2 then changes from a superposition to a classical scenario, at which the photon is either completely reflected by or passes the beam splitter, as shown in the insets.

After the spacetime border has passed the critical expansion ($\bar{t} > \bar{t}_C$), the evolution of the wavefunction remains a classical scenario, since according to our stability analysis, intensity shifts back to superpositions are only possible for $\bar{t} = \bar{t}_C$.

### 4.3. Lifetimes of superpositions in agreement with the Diósi-Penrose criterion

Based on the result in (37) for the time instance of collapse, we can estimate the lifetimes of the superpositions. For the case where the Diósi-Penrose energy in our experiment $E_{DP\,12}(t)$ reaches a constant value of $E_{DP\,12}$ shortly after the superposition has split at $t_s$, we can use (37), (29) and set the splitting point of the superposition to $t_s = 0$ to get:

$$\bar{t}_C \approx \frac{\hbar}{E_{DP\,12}}, \qquad (38)$$

which agrees with the Diósi-Penrose criterion (16).

### 4.4. No statistical variation in the reduction point in time

An interesting feature of our collapse model is that the reduction point in time $\bar{t}_C$ does not vary statistically over multiple executions of the same experiment (except, of course, for variations which have their origin in statistical variations of the experimental parameters). This follows from the collapse condition in (37), according to which collapse is only possible when the competition action between the paths is exactly $\hbar$.

### 4.5. Energy conservation

Since the evolution of the wavefunction changes after collapse to classical scenarios, which in our experiment are identical to cases where the reflection coefficient of the beam splitter is set to 100% or 0%, the energies for these classical scenarios are expected to be constant over time. Hence, the energy after collapse is identical to the energy before the superposition has split. It is only during superposition that the frequencies of the wavefunctions for the paths are slightly lower than expected from the unitary evolution, as discussed in Section 3.4.

### 4.6. Quantum correlations

With the collapse model developed thus far for two-state superpositions, we can explain the quantum correlations occurring in single photon and EPR experiments.

In the single photon experiment shown in the upper part of Figure 4, in which either the left or the right detector detects the photon (but never both together), the evolution of the wavefunction will change at collapse to the classical scenario in which the photon is either completely reflected by or passes through the beam splitter. This explains why the photon can never be detected by both detectors, independent of how far the detectors are separated from each other.

To apply our collapse model for two-state superpositions to the EPR experiment in the lower part of Figure 4, in which Alice and Bob can freely choose the orientation of their polarisation filters shortly before the arrival of the photon, we simplify the discussion by choosing Alice's arm to be at least $c(\bar{t}_C - t_s)$ longer than Bob's, which ensures that the superposition is reduced by Bob's measurement. Bob can then determine the two polarisation states between which a competition action is built up by choosing the orientation of his polarisation filter. When this competition action (caused by Bob's measurement) reaches $\hbar$, the superposition is reduced to one of the polarisation states that Bob has determined by his choice. Since the abrupt change in the evolution of the wavefunction covers both Bob's and Alice's locations, the polarisation of Alice's photon changes instantaneously (based on Bob's choice) before it arrives at her detector. Thus, Alice will observe the correlations predicted by quantum theory for any choice of the orientation of her polarisation filter.

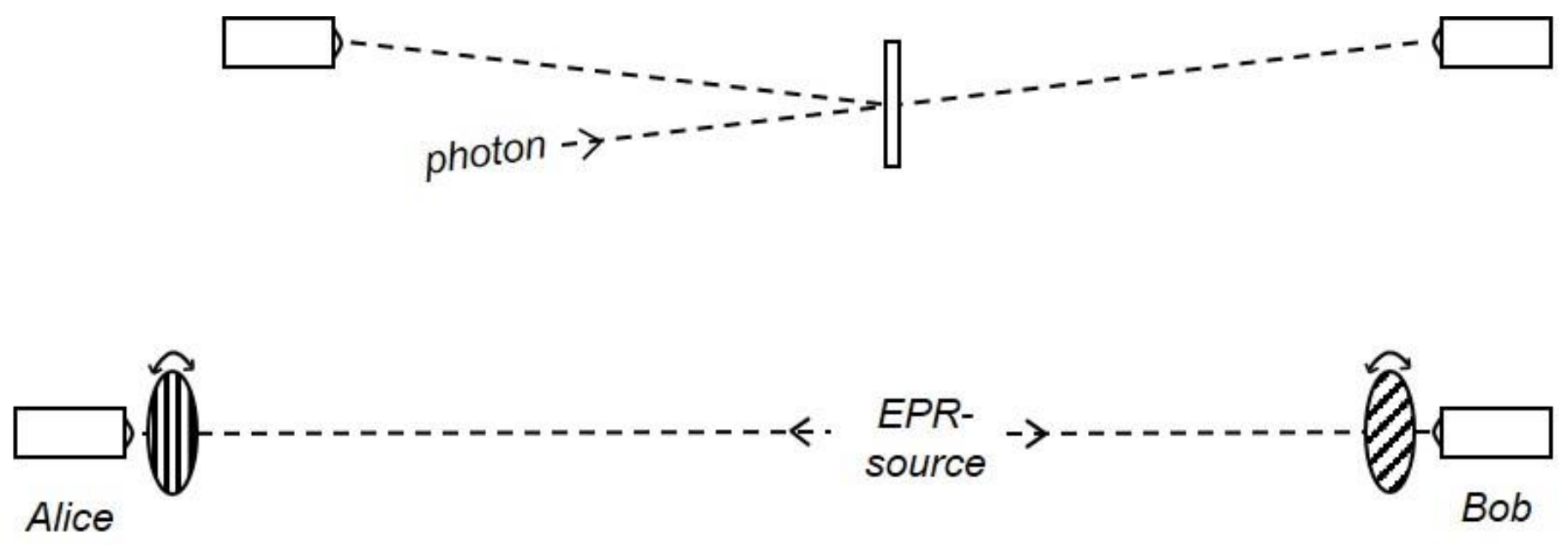

**Figure 4.** Experiments used in the discussion of quantum correlations. *Upper*: Single-photon experiment in which either the left or the right detector measures the photon. *Lower*: EPR experiment with a free choice of measurement, in which Alice and Bob can freely choose the orientation of their polarisation filters shortly before the arrival of the photon.

This explanation of quantum correlations does not come into conflict with relativity, since the evolution of the wavefunction before and after collapse is governed solely by local physical laws, in agreement with relativity.

# 5. Reduction probabilities and Born's rule

In this section, we will show that the Temporally Expanding Spacetime approach naturally leads to stochastic behaviour, that it can predict reduction probabilities, and that it reproduces Born's rule for two-state superpositions. The Temporally Expanding Spacetime approach is the first approach to collapse to make reduction probabilities a subject of prediction. In the collapse models developed thus far, reduction probabilities are not an issue, presumably because there are many proofs of Born's rule that derive it from simple, basic assumptions [46-53], and because exceptions from Born's rule can lead to faster-than-light signalling, as we shall see in Section 9. The proofs of Born's rule deriving it from basic assumptions cannot be applied to our collapse model, since they consider the measurement process very generally without specifying the measurement apparatus used in the individual case, as in our model.

## 5.1. Decay-trigger rates

The probability with which the evolution of the wavefunction collapses at the critical spacetime expansion $\bar{t}_C$ to one of the classical scenarios can be derived from the following physical argument. According to the discussion in Section 4.2, fluctuations with the smallest intensity in favour of one of the paths are sufficient to induce collapse of the wavefunction's evolution at the critical spacetime expansion $\bar{t}_C$. In our model, such fluctuations follow in a natural way from the permanent expansion of spacetime. According to (27) and (29), the competition actions of paths 1 and 2 permanently increase as the spacetime border moves towards the future, and are $dS_{det\,1}$=$I_2 E_{DP\,12}(\bar{t})d\bar{t}$ and $dS_{det\,2}$=$I_1 E_{DP\,12}(\bar{t})d\bar{t}$ when the spacetime border has moved by $d\bar{t}$. According to the discussion in Section 4.1, a path tries to react to an increase in its detuning action with an intensity shift in favour of the other path (cf. the equations in (34)), which we refer to as a *decay trigger* for the path. These decay triggers (resulting from the permanent increase in the detuning actions of the paths) will induce collapse of the wavefunction's evolution at the critical spacetime expansion $\bar{t}_C$. The probabilities with which the wavefunction's evolution collapses at $\bar{t}_C$ to either classical scenario 1 or 2 depends on how often the decay triggers cause decay of the paths. We therefore define *decay-trigger rates* $\nu_{\downarrow trig}$ for the paths, which describe how often the decay triggers cause decay of the path over the spacetime expansion parameter $\bar{t}$. The decay-trigger rates of the paths are expected to be proportional to the increase in the detuning actions of the paths over $\bar{t}$:

$$\nu_{\downarrow trig\, i}(\bar{t}) \sim \frac{d}{d\bar{t}} S_{det\, i}(\bar{t}) . \tag{39}$$

From (27) and (29), we obtain the following decay-trigger rates for paths 1 and 2:

$$\begin{aligned} \nu_{\downarrow trig\, 1}(\bar{t}) &\sim \frac{d}{d\bar{t}} S_{det\, 1}(\bar{t}) = I_2\, E_{DP\, 12}(\bar{t}) \\ \nu_{\downarrow trig\, 2}(\bar{t}) &\sim \frac{d}{d\bar{t}} S_{det\, 2}(\bar{t}) = I_1\, E_{DP\, 12}(\bar{t}) \end{aligned} . \tag{40}$$

## 5.2. Born's rule

Since at collapse, path 1 decays in favour of path 2, and vice versa, the reduction probability for classical scenario 1 $p_1$ is proportional to the decay-trigger rate of path 2, and similarly for the reduction probability for classical scenario 2 $p_2$:

$$\begin{aligned} p_1 &\sim \nu_{\downarrow trig\, 2}(\bar{t}_C) \\ p_2 &\sim \nu_{\downarrow trig\, 1}(\bar{t}_C) \end{aligned} . \tag{41}$$

Since the decay-trigger rate for path 2 is proportional to the intensity of path 1 ($\nu_{\downarrow trig\, 2} \sim I_1$), in favour of which it will decay at collapse and vice versa ($\nu_{\downarrow trig\, 1} \sim I_2$), we get reduction probabilities for the classical scenarios that are proportional to their intensities ($p_1 \sim I_1$, $p_2 \sim I_2$), which leads to Born's rule:

$$\begin{aligned} p_1 &= I_1 \\ p_2 &= I_2 \end{aligned} . \tag{42}$$

## 5.3. Decay-trigger amplitudes

An important feature of our collapse model is that fluctuations with the smallest intensity are sufficient to induce the self-reinforcing loop at the critical spacetime expansion $\bar{t}_C$, thus inducing collapse (cf. Section 4.2). This means that only the frequencies of the decay triggers, rather than their amplitudes, have an impact on the reduction probabilities. From Equation (33), which describes how much intensity $dI$ is attempted to be shifted to the other path when the detuning action of the path is increased by $dS_{det}$, which is a measure for the amplitude of the decay-trigger, it follows that the decay-trigger amplitudes $a_{\downarrow trig}$ are proportional to the intensity $I$ of the path, i.e.:

$$a_{\downarrow trig\, i} \sim I_i . \tag{43}$$

We will come back to this result in Section 9 when discussing the exceptions to Born's rule.

# 6. Covariant formulation

In this section, we will formulate our collapse model in a covariant way and show how the competition action is related to the Einstein-Hilbert action. Furthermore, for the evolution of the spacetime border, we consider the general case in which $\bar{\sigma}(\bar{\tau})$ may be any sequence of time-ordered spacelike hypersurfaces, and must not be aligned with the rest frame of the experiment. Since we return in Section 7 to the Newtonian limit and to this assumption, the study of this section is optional.

## 6.1. Covariant formulation of the wavefunction's evolution

In the relativistic equivalent to Schrödinger's equation, known as the Tomonaga-Schwinger equation, the evolution of the state vector $|\psi(\tau)>$ can be followed up for arbitrary sequences of time-ordered spacelike hypersurfaces $\sigma(\tau)$ [54,55], where the sequence parameter $\tau$ can be chosen to be dimensionless, as the spacetime expansion parameter $\bar{\tau}$. The evolution of the state vector from $|\psi(\tau)>$ on the hypersurface $\sigma(\tau)$) to $|\psi(\tau + d\tau)>$ on $\sigma(\tau + d\tau)$ is determined by an integral over a Hamilton density on the spacetime volume between $\sigma(\tau)$ and $\sigma(\tau + d\tau)$ [56].

Using the path-integral method, the amplitude $\psi_{conf}$ of e.g. a configuration of particles on $\sigma(\tau)$ is given by a sum over all paths leading to this configuration on $\sigma(\tau)$: $\psi_{conf} \sim \sum_{path} e^{\frac{i}{\hbar}\int d\tau L_{path}(\tau)}$, where $L_{path}(\tau)$ is the Lagrange function of a path over the sequence parameter $\tau$. For both methods, the calculation of the state vector on a hypersurface $\sigma_2$ from the state vector on a preceding hypersurface $\sigma_1$ is independent of the hypersurface sequence $\sigma(\tau)$, which is chosen between $\sigma_1$ and $\sigma_2$ for following up the evolution, which represents the Lorentz invariance.

In classical scenarios, the amplitude of the wavefunction along the classical trajectories of the particles $\psi_{cl\,sc}$ can be calculated based on the argument of the stationary phase, which yields $\psi_{cl\,sc}(\tau) \sim e^{\frac{i}{\hbar}\int_{..}^{\tau} d\tau L_{cl\,sc}(\tau)}$, where $L_{cl\,sc}(\tau)$ is the Lagrange function of the particles along their classical trajectories on $\sigma(\tau)$. For a free particle with mass $m$, the Lagrange function is: $\int_{..}^{\tau} d\tau' L_{cl\,sc}(\tau') = -\int ds\, mc$ [41]. For an ensemble of particles, this result can be generalised to $-\int_{..}^{\sigma(\tau)} \frac{d^4x}{c}\sqrt{-g_{cl\,sc}(x)}\, T_{cl\,sc}(x)$. Here, $T_{cl\,sc}(x)$ is the contracted energy momentum tensor field of the particle ensemble ($T(x) = T^{\mu}{}_{\mu}(x)$), and $\sqrt{-g_{cl\,sc}(x)}$ is the covariant volume element ($\sqrt{-g} = \sqrt{-\det(g_{\mu\nu})}$), where the metric field $g_{\mu\nu\,cl\,sc}(x)$ for the calculation of $\sqrt{-g_{cl\,sc}(x)}$ follows from the energy momentum tensor field of the particle ensemble $T_{\mu\nu\,cl\,sc}(x)$ with Einstein's field equations. The amplitude of the wavefunction along the classical trajectory is then given by:

$$\psi_{cl\,scen}(\tau) \sim e^{-\frac{i}{\hbar}\int_{..}^{\sigma(\tau)} \frac{d^4x}{c}\sqrt{-g_{cl\,sc}(x)}\, T_{cl\,sc}(x)}\,. \quad (44)$$

### 6.2. Covariant formulation of the collapse model

In the Newtonian limit, the contracted energy momentum tensor field $T(x)$ and the covariant volume element $\sqrt{-g(x)}$ in Equation (44) become [41]:

$$\begin{aligned} T(x) &\to \rho(\mathbf{x},t)c^2 \\ \sqrt{-g(x)} &\to 1+\frac{\Phi(\mathbf{x},t)}{c^2} \ . \end{aligned} \tag{45}$$

With these replacements, (44) can be simplified to:

$$\psi_{cl\,sc}(t) \sim e^{-\frac{i}{\hbar}\int_{..}^{t} dt \int d^3\mathbf{x}\rho_{cl\,sc}(\mathbf{x},t)(c^2+\Phi_{cl\,sc}(\mathbf{x},t))}\ , \tag{46}$$

which is the result expected for the Newtonian limit. In the following, we will show that the formulas derived thus far can be transformed into covariant form by replacing $\rho_i(\mathbf{x},t)$ by $T_i(x)/c^2$ and $\Phi_i(\mathbf{x},t)$ by $c^2(\sqrt{-g_i(x)}-1)$, which is suggested from the relations in (45).

The energy momentum tensor field $T_{\mu\nu}(x)$ of a two-state superposition ($|\psi>= c_1|\psi_1> + c_2|\psi_2>$) can be calculated in an analogous way to the mass density in (5), with help of an operator $\hat{T}_{\mu\nu}(x)$, which yields:

$$T_{\mu\nu}(x) = |c_1|^2 T_{\mu\nu\,1}(x) + |c_2|^2 T_{\mu\nu\,2}(x)\ , \tag{47}$$

with $T_{\mu\nu\,i}(x) =< \psi_i|\hat{T}_{\mu\nu}(x)\,|\psi_i >$.

In our calculations, we refer to the limit of weak gravitational fields, for which Einstein's field equations can be linearised, and for which the metric field $g_{\mu\nu}(x)$ deviates only slightly from Minkowski' s metric $\eta_{\mu\nu}$ ($\eta_{\mu\nu} \equiv \mathrm{diag}(1,-1,-1,-1)$) [41]:

$$g_{\mu\nu}(x) = \eta_{\mu\nu} + h_{\mu\nu}(x)\ ; \quad |h_{\mu\nu}(x)| << 1\ . \tag{48}$$

In this linear regime, we have linear mappings from the energy momentum tensor field $T_{\mu\nu}(x)$ to the $h_{\mu\nu}(x)$-field and to the difference of the covariant volume element from one ($\sqrt{-g(x)}-1$). From this and the relation in (47), it follows that ($\sqrt{-g(x)}-1$) can be decomposed into:

$$\sqrt{-g_{com}(x)} - 1 = |c_1|^2\left(\sqrt{-g_1(x)}-1\right) + |c_2|^2\left(\sqrt{-g_2(x)}-1\right), \tag{49}$$

where $\sqrt{-g_i(x)}$ is the covariant volume element of path $i$ resulting from its metric field $g_{\mu\nu\,i}(x)$, which follows from its energy momentum tensor field $T_{\mu\nu\,i}(x)$ with Einstein's field equations.

To calculate the phase changes for the paths at the splitting point of the superposition in (20), in the case where the evolution of the spacetime border $\bar{\sigma}(\bar{\tau})$ may be any sequence of time-ordered spacelike hypersurfaces, we choose a special hypersurface sequence for following up the wavefunction's evolution $\sigma_{\bar{\tau}}(\tau)$, which matches the spacetime border at $\tau = \bar{\tau}$ (i.e. $\sigma_{\bar{\tau}}(\bar{\tau}) = \bar{\sigma}(\bar{\tau})$) and touches the splitting point of the superposition at $\tau = \tau_s$. The phase changes in the wavefunction of the paths $\Delta\varphi_i$, which occur when intensity between the paths is shifted and which led us in Section 3.4 to the result in (20), are given for the covariant case by (cf. the relations in (45)):

$$\hbar\Delta\varphi_i = -\int_{\sigma_{\bar{\tau}}(\tau_s)}^{\sigma_{\bar{\tau}}(\bar{\tau})} \frac{d^4x}{c} T_i(x)\Delta\sqrt{-g_{com}(x)}\,, \tag{50}$$

where $\Delta\sqrt{-g_{com}(x)}$ is the change in the covariant volume element resulting from the intensity shift $dI$ between the paths (cf. Section 3.4). From the results in (25), (24) and (23), it follows that the detuning action of path 1 is related to the phase changes by $S_{det\,1} = \hbar(\Delta\varphi_2 - \Delta\varphi_1)/2$, which leads with (50) to:

$$S_{det\,1}(\bar{\tau}) = -\frac{1}{2}\int_{\sigma_{\bar{\tau}}(\tau_s)}^{\sigma_{\bar{\tau}}(\bar{\tau})} \frac{d^4x}{c}(T_2(x) - T_1(x))\Delta\sqrt{-g_{com}(x)}\ . \tag{51}$$

The change in the covariant volume element $\Delta\sqrt{-g_{com}(x)}$ when the intensity of path 1 is decreased from one to $|c_1|^2$ to calculate its absolute frequency detuning (cf. Section 3.4) follows from (49), and yields with $dI = |c_2|^2$:

$$\Delta\sqrt{-g_{com}(x)} = |c_2|^2\left(\sqrt{-g_2(x)} - \sqrt{-g_1(x)}\right). \tag{52}$$

By inserting this into (51) and calculating the detuning action of path 2 in an analogous way to that of path 1, we get the following detuning actions for path 1 and 2 (cf. (27)):

$$\begin{aligned} S_{det\,1}(\bar{\tau}) &= |c_2|^2\, S_{comp\,12}(\bar{\tau}) \\ S_{det\,2}(\bar{\tau}) &= |c_1|^2\, S_{comp\,12}(\bar{\tau})\,' \end{aligned} \tag{53}$$

where the competition action $S_{comp\,12}(\bar{\tau})$ is given by:

$$S_{comp\,12}(\bar{\tau}) = -\frac{1}{2}\int_{..}^{\bar{\sigma}(\bar{\tau})} \frac{d^4x}{c}\big(T_2(x) - T_1(x)\big)\left(\sqrt{-g_2(x)} - \sqrt{-g_1(x)}\right), \tag{54}$$

which is the expected covariant generalisation of our former result in (30).

The stability analysis of the evolution of the wavefunction in Section 4.1 can be adapted for the covariant case by simply replacing the competition action $S_{comp\,12}(\bar{t})$ in the formulae by the generalised version $S_{comp\,12}(\bar{\tau})$, which leads to the following reduction condition:

$$S_{comp\,12}(\bar{\tau}_C) = \hbar\,. \tag{55}$$

In the derivation of Born's rule given in Section 5, the definition of the decay-trigger rates (39) is generalised to:

$$\nu_{\downarrow trig\,i}(\bar{\tau})\ \sim \frac{d}{d\bar{\tau}} S_{det\,i}(\bar{\tau})\,, \tag{56}$$

which with (53) leads to the following decay-trigger rates for paths 1 and 2:

$$\begin{aligned}\nu_{\downarrow trig\,1}(\bar{\tau}) &\sim \tfrac{d}{d\bar{\tau}} S_{det\,1}(\bar{\tau}) = |c_2|^2 \tfrac{d}{d\bar{\tau}} S_{comp\,12}(\bar{\tau}) \\ \nu_{\downarrow trig\,2}(\bar{\tau}) &\sim \tfrac{d}{d\bar{\tau}} S_{det\,2}(\bar{\tau}) = |c_1|^2 \tfrac{d}{d\bar{\tau}} S_{comp\,12}(\bar{\tau})\end{aligned} \,. \tag{57}$$

As the result in (40), we see that these decay-trigger rates are proportional to the intensity of the competing path in favour of which the path decays at collapse, hence leading to Born' s rule.

### 6.3. Competition and Einstein-Hilbert action

In this section, we show how the covariant generalisation of the competition action (54) is related to the fundamental action of general relativity, the Einstein-Hilbert action. The Einstein-Hilbert action on the spacetime region up to the spacetime border $\bar{\sigma}(\bar{\tau})$ is given by [41]:

$$S_{EH}(\bar{\tau}) = \int_{..}^{\bar{\sigma}(\bar{\tau})} \frac{d^4x}{c} \sqrt{-g(x)}\, \frac{R(x)}{2\kappa} \,, \tag{58}$$

where $R(x)$ is the tension scalar, and $\kappa$ is defined by: $\kappa \equiv 8\pi G/c^4$. To derive the relation between the competition and the Einstein-Hilbert action, we calculate the Einstein-Hilbert action for a superposition. To do this, we express the tension scalar $R(x)$ in (58) by the contracted energy momentum tensor field $T(x)$ with the help of the relation $R(x) = \kappa T(x)$, which follows from a contraction of Einstein's field equations [41]. With $T(x) = |c_1|^2 T_1(x) + |c_2|^2 T_2(x)$ (following from (47)) and $\sqrt{-g_{com}(x)} = |c_1|^2\sqrt{-g_1(x)} + |c_2|^2\sqrt{-g_2(x)}$ (following from (49) and the norm conservation), and after a short calculation,[2] we get:

$$S_{EH}(\bar{\tau}) = |c_1|^2 S_{EH\,1}(\bar{\tau}) + |c_2|^2 S_{EH\,2}(\bar{\tau}) + |c_1|^2|c_2|^2 S_{comp\,12}(\bar{\tau}) \,, \tag{59}$$

where $S_{EH\,1}(\bar{\tau})$ and $S_{EH\,2}(\bar{\tau})$ are the Einstein-Hilbert actions corresponding to paths 1 and 2. This result can be interpreted in an analogous way to the expression in (17) for the gravitational self-energy of a superposition. The term $|c_1|^2|c_2|^2 S_{comp\,12}(\bar{\tau})$ in (59) results from the fact that the paths do not reside in their preferred spacetime geometries (resulting from their mass distributions), but must share a common spacetime geometry under semiclassical gravity.

### 6.4. Lorentz invariance

It is obvious that our collapse model is not Lorentz invariant in the ordinary sense, since the spacetime border distinguishes one Lorentz frame from the others.

---

[2] $(|c_1|^2 T_1 + |c_2|^2 T_2)(|c_1|^2\sqrt{-g_1} + |c_2|^2\sqrt{-g_2}) =$
$|c_2|^2 T_2(|c_1|^2\sqrt{-g_1} + |c_2|^2\sqrt{-g_2}) + |c_1|^2 T_1(|c_1|^2\sqrt{-g_1} + |c_2|^2\sqrt{-g_2}) =$
$|c_2|^2 T_2(\sqrt{-g_2} - |c_1|^2\sqrt{-g_2} + |c_1|^2\sqrt{-g_1}) + |c_1|^2 T_1(\sqrt{-g_1} - |c_2|^2\sqrt{-g_1} + |c_2|^2\sqrt{-g_2}) =$
$|c_1|^2\sqrt{-g_1}T_1 + |c_2|^2\sqrt{-g_2}T_2 - |c_1|^2|c_2|^2(T_2 - T_1)(\sqrt{-g_2} - \sqrt{-g_1}).$

However, our model can be considered to be relativistic in the sense that in the spacetime region before the spacetime border, we are free to choose the hypersurface sequences to follow up the evolution of the state vector, and that our collapse model can be formulated in a covariant way.

Since the covariant formulation of our collapse model has no advantages in terms of the discussion of concrete experiments, we return in the following sections to our former assumptions, the Newtonian limit, and propagation of the spacetime border that is aligned with the reference frame of the experiment. However, all of the results derived in the following can easily be transformed to covariant form using the expressions derived here.

# 7. Superpositions of more than two states

In this and the following section, we will address the question of whether the two most important results of our collapse model, the explanation of the wavefunction collapse by abrupt changes in the evolution of the wavefunction, and the reproduction of Born's rule, also follow for superpositions of more than two states. In this section, the collapse model is extended to handle superpositions of more than two states, and in Section 8 it is applied to typical quantum mechanical experiments.

## 7.1. Local bundles, Diósi-Penrose energies, and competition actions

The following extension of our collapse model to superpositions of more than two states involves an approximation that is well satisfied for all experiments performed so far. This approximation will be explained based on the three-detector experiment in Figure 5, in which a photon is split into three beams and measured in each beam with the same type of detector as in Figure 2 (shifting a rigid body when a photon is measured). The state vector describing detector 1 in this experiment (i.e. the state vector describing the particles belonging to this detector) can only distinguish the cases in which the photon is measured by this detector or not. Cases in which the photon is detected by detectors 2 or 3 cannot be distinguished. Hence, state vectors describing the detection of the photon by detectors 2 or 3 can be decomposed into a product of a factor referring to the location of detector 1 $D1$ and a factor referring to the area outside of $D1$, as: $|\psi_2 >= |\psi >_{D1} \otimes |\psi_2 >_{\neg D1}$, $|\psi_3 >= |\psi >_{D1} \otimes |\psi_3 >_{\neg D1}$. When the gravitational potentials resulting from the mass distributions of states 2 and 3, $\Phi_2(\mathbf{x})$ and $\Phi_3(\mathbf{x})$, are almost identical in the area of detector 1 $D1$, we combine states 2 and 3 to form a so-called *local bundle* $b_{(2,3)}^{D1}$ on $D1$, where $D1$ is known as the *bundle area*.

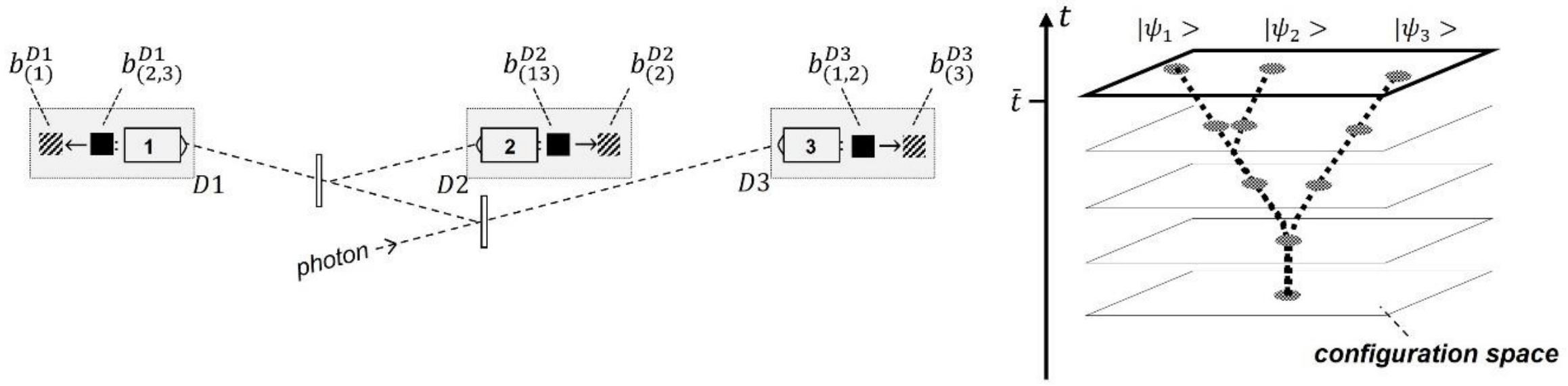

**Figure 5.** *Left:* Three-detector experiment in which a single photon is split into three beams and measured in each beam by a detector as in Fig. 2 (shifting a rigid body when a photon is measured). *Right:* Evolution of the state vector of the experiment in configuration space.

In our three-detector experiment, we have three bundle areas, $D1$, $D2$ and $D3$, which refer to the locations of the three detectors (see Figure 5). For each bundle area, we have two local bundles, which refer to the cases of "detection" and "no detection". For detector 1, these are denoted as $b_{(1)}^{D1}$ and $b_{(2,3)}^{D1}$, and accordingly for the other detectors (see Figure 5). It is important to realise that each state of a superposition is an element of exactly one local bundle on each bundle area, which can easily be understood based on our experiment in Figure 5. The local bundles that can be defined for a bundle area must not consist of several states. The three local bundles in Figure 5 that refer to cases of "detection", which are $b_{(1)}^{D1}$, $b_{(2)}^{D2}$, and $b_{(3)}^{D3}$, consist of only one state.

For a local bundle $b_\kappa^A$ on a bundle area $A$, we can define a bundle intensity $I_\kappa^A$ by the sum of the intensities of all states $i$ belonging to this bundle:

$$I_\kappa^A \equiv \sum_{i \in b_\kappa^A} I_i \,. \tag{60}$$

The sum over the intensities $I_\kappa^A$ of all local bundles $b_\kappa^A$ belonging to a bundle area $A$ is one ($\sum_\kappa I_\kappa^A = 1$).

Between two local bundles $b_\kappa^A$ and $b_\nu^A$ that refer to the same bundle area $A$, we can define the so-called *local Diósi-Penrose energies* $E_{DP\,\kappa\nu}^A$ and *local competition actions* $S_{comp\,\kappa\nu}^A(\bar{t})$ as follows. Since the states belonging to a local bundle have identical state vectors and mass distributions, and almost identical gravitational potentials on the bundle area, we can define the local Diósi-Penrose energies between the local bundles $b_\kappa^A$ and $b_\nu^A$ using (13), as follows:

$$E_{DP\,\kappa\nu}^A \equiv -\frac{1}{2} \int_{\mathbf{x} \in A} d^3\mathbf{x}\, (\rho_\nu^A(\mathbf{x}) - \rho_\kappa^A(\mathbf{x}))(\Phi_\nu^A(\mathbf{x}) - \Phi_\kappa^A(\mathbf{x}))\,, \tag{61}$$

where $\rho_\nu^A(\mathbf{x})$, $\rho_\kappa^A(\mathbf{x})$, $\Phi_\nu^A(\mathbf{x})$ and $\Phi_\kappa^A(\mathbf{x})$ are the mass densities and gravitational potentials of the local bundles, respectively. Based on the result in (29), we can use the local Diósi-Penrose energy $E_{DP\,\kappa\nu}^A(t)$ to calculate a local competition action $S_{comp\,\kappa\nu}^A(\bar{t})$ between $b_\kappa^A$ and $b_\nu^A$ as:

$$S_{comp\,\kappa\nu}^A(\bar{t}) \equiv \int_{..}^{\bar{t}} dt\, E_{DP\,\kappa\nu}^A(t)\,. \tag{62}$$

### 7.2. Conditional equation for wavefunction collapse

To generalise the conditional equation for the collapse of two-state superpositions (35) to superpositions of more than two states, we first examine the case in which only the local Diósi-Penrose energy between two local bundles $b_\kappa^A$ and $b_\nu^A$ on one bundle area $A$ is non-vanishing. In this case, the conditional equation for the intensity changes $dI_\kappa^A$, $dI_\nu^A$ of the two competing local bundles $b_\kappa^A$, $b_\nu^A$ must be identical to the equation in (35):

$$\begin{aligned} dI_\kappa^A &= \frac{S_{comp\ \kappa\nu}^A(\bar{t})}{\hbar}\left(I_\nu^A dI_\kappa^A - I_\kappa^A dI_\nu^A\right) \\ dI_\nu^A &= \frac{S_{comp\ \kappa\nu}^A(\bar{t})}{\hbar}\left(I_\kappa^A dI_\nu^A - I_\nu^A dI_\kappa^A\right) \end{aligned} . \tag{63}$$

This result can be extended to the general case as follows. The final conditional equation is a set of $N$ equations for the intensity changes $dI_i$ for each path of the superposition. For a path $i$, we determine with the function $\kappa(A,i)$ for each bundle area $A$ the local bundle $b_\kappa^A$ that contains this path ($i \in b_\kappa^A$). The intensity changes of the local bundles $dI_\kappa^A$ in (63) are transformed into intensity changes of the paths $dI_i$ with help of the projections $dI_i = (dI_i/dI_\kappa^A)dI_\kappa^A$. In addition, we must introduce one sum over all of the local bundles $b_\nu^A$ on $A$ that are competing with the considered bundle $b_\kappa^A$ on $A$, and a further sum over all bundle areas $A$. This leads to the following conditional equation for the intensity changes $dI_i$ of the paths:

$$dI_i = \sum_A \frac{dI_i}{dI_\kappa^A} \sum_{\nu \neq \kappa(A,i)} \frac{S_{comp\ \kappa\nu}^A(\bar{t})}{\hbar}\left(I_\nu^A dI_\kappa^A - I_\kappa^A dI_\nu^A\right) , \tag{64}$$

which we will apply in Sections 8 and 9 to concrete experiments. For our three-detector experiment in Figure 5, the outer sum of this equation runs over the three bundle areas $D1$, $D2$ and $D3$. The inner sum consists of only one element, since at the locations of the three detectors, we have only two competing bundles, which refer to the cases of “detection” and “no detection”.

### 7.3. Final states after collapse

In this section, we will calculate the final states after collapse and show that the key feature of our collapse model, i.e. the abrupt changes in the evolution of the wavefunction at the critical expansions of spacetime $\bar{t}_C$, also follows for superpositions of more than two states. In the following discussion, we abbreviate the intensities of the paths of superposition $I_i$ by an intensity vector $\vec{I}$ :

$$\vec{I} \equiv \begin{pmatrix} I_1 \\ I_2 \\ \cdot\cdot \\ I_N \end{pmatrix} . \tag{65}$$

If for a given intensity vector $\vec{I}$ we have found a critical spacetime expansion $\bar{t}_C$ with the solution $d\vec{I}_C \equiv (dI_{1C}, dI_{2C}, ..)^T$ for the conditional equation in (64), it follows that $d\vec{I}_C$ is then also a solution for $\vec{I}' = \vec{I} \pm \alpha\, d\vec{I}_C$, which directly follows by inserting $\vec{I}'$ into (64). When the intensity vector $\vec{I}$ is fluctuating at $\bar{t} = \bar{t}_C$ in the $+d\vec{I}_C$-direction, the intensity vector $\vec{I}$ can then change along the line: $\vec{I}' = \vec{I} + \alpha\, d\vec{I}_C$. This change is expected to happen abruptly, since the initial intensity fluctuation of $\vec{I}$ leads to a change in spacetime geometry, which reinforces this fluctuation. Thus, we are entering the self-reinforcing loop described in Section 4.2 for the collapse of a two-state superposition. This change in the intensity vector will not stop until one of its components has reached zero. The final intensity vector of this process is denoted as $\vec{I}'_+$ and the corresponding value of $\alpha$ as $\alpha_+$ ($\vec{I}'_+ = \vec{I} + \alpha_+\, d\vec{I}_C$). In the same way, when the intensity vector $\vec{I}$ is fluctuating at $\bar{t} = \bar{t}_C$ in the $-d\vec{I}_C$-direction, the intensity vector will abruptly change along the line $\vec{I}' = \vec{I} - \alpha\, d\vec{I}_C$ to $\vec{I}'_- = \vec{I} - \alpha_-\, d\vec{I}_C$, where $\alpha_-$ is defined accordingly. In Sections 8 and 9, we will apply this procedure to calculate the final states after collapse for concrete experiments.

### 7.4. Reduction probabilities

Finally, we must calculate the reduction probabilities for the final states after collapse in the previous section. In the derivation of Born's rule for two-state superpositions in Section 5, we determined decay-trigger rates for the paths based on the increase in their detuning actions over $\bar{t}$ (cf. the equation in (40)). Here, we determine a detuning action for each local bundle $b_\kappa^A$ on each bundle area $A$, and use this to calculate a corresponding decay-trigger rate for the bundle.

The detuning action $S_{det\,\kappa}^A$ of a local bundle $b_\kappa^A$ can be calculated as follows. When only the competition action between two local bundles $b_\kappa^A$ and $b_\nu^A$ on one bundle area $A$ is non-vanishing, we expect detuning actions for these bundles in the same way as for the two states of a two-state superposition (cf. the equation in (27)):

$$\begin{aligned} S_{det\,\kappa}^A(\bar{t}) &= I_\nu^A\, S_{comp\,\kappa\nu}^A(\bar{t}) \\ S_{det\,\nu}^A(\bar{t}) &= I_\kappa^A\, S_{comp\,\kappa\nu}^A(\bar{t}) \end{aligned} . \qquad (66)$$

When we have more than two local bundles on $A$, we must take into account when calculating the detuning actions in Section 3.4 that when intensity is shifted away from the considered local bundle, this intensity is divided among the other local bundles on $A$. This leads to the following generalisation for the detuning action of $b_\kappa^A$ [3]:

$$S_{det\,\kappa}^A(\bar{t}) = \sum_{\nu \neq \kappa} I_\nu^A\, S_{comp\,\kappa\nu}^A(\bar{t}) . \qquad (67)$$

---

[3] To ensure that this result is independent of whether the gravitational self-energies $E_{\mathrm{G}\,self\,i}$ of the superposed states are different (as shown in Section 3.4 for two-state superpositions), we need to calculate the phase shift $\Delta\varphi_k$ of a path $k$ at the splitting point of the superposition $t_s$ when intensity is shifted between any two paths $i$ and $j$ ($I_i \to I_i - dI,\ I_j \to I_j + dI$). In this calculation, result (22) generalises to:

$$\hbar\Delta\varphi_k = -dI \int_{t_s}^{\bar{t}} dt\, \left(E_{DP\,jk}(t) - E_{DP\,ik}(t) + E_{\mathrm{G}\,self\,j}(t) - E_{\mathrm{G}\,self\,i}(t)\right),$$

where $k$ can be $k = i, j$ or $k \neq i, j$. This shows that different gravitational self-energies lead to the same phase shift $\Delta\varphi_k$ for all paths, and thus have no influence on the frequency detuning of the paths.

The decay-trigger rates of the local bundles $b_\kappa^A$ can be defined in an analogous way to the decay-trigger rates of the paths in (39), as follows:

$$v_{\downarrow trig\ \kappa}^A(\bar{t}) \sim \frac{d}{d\bar{t}} S_{det\ \kappa}^A(\bar{t})\,. \qquad (68)$$

When calculating the reduction probabilities for the final states after collapse, $\vec{I}'_+$ and $\vec{I}'_-$, we must determine whether the decay-trigger rates of the local bundles allow the intensity vector to fluctuate in the $+d\vec{I}_C$- or $-d\vec{I}_C$-direction. When the intensity of a local bundle $b_\kappa^A$ decreases as the intensity vector is shifted in the $+d\vec{I}_C$- direction ($I_\kappa^A(\vec{I} + d\vec{I}_C) < I_\kappa^A(\vec{I})$, its decay-trigger rate triggers a decay of the superposition in favour of $\vec{I}'_+$. In the same way, a decay in favour of $\vec{I}'_-$ is triggered when the bundle's intensity decreases as the intensity vector is shifted in the $-d\vec{I}_C$- direction ($I_\kappa^A(\vec{I} - d\vec{I}_C) < I_\kappa^A(\vec{I})$. By summing up all decay-trigger rates triggering a decay in favour of $\vec{I}'_+$ or $\vec{I}'_-$ we get the reduction probabilities of these states. We will demonstrate the application of this procedure to concrete experiments in the following sections.

# 8. Typical quantum mechanical experiments

In this section, we apply our collapse model for superpositions of more than two states to typical quantum mechanical experiments.

## 8.1. A common property of all experiments performed so far

Typical quantum mechanical experiments can be categorised into two groups: experiments with active measuring devices, such as the three-detector experiment in Figure 5, and experiments with passive measuring devices, for example films or cloud chambers. Both groups can be discussed as one by modelling passive measuring devices with many small mass-displacing detectors, as shown in Figure 6 for a photon measurement with a film. The common property of all these experiments is that at the locations of the detectors, we never have more than two local bundles, which refer to "detection" and "no detection" by the detector. Based on this property, we derive Born's rule and show how the particles are localised during measurements.

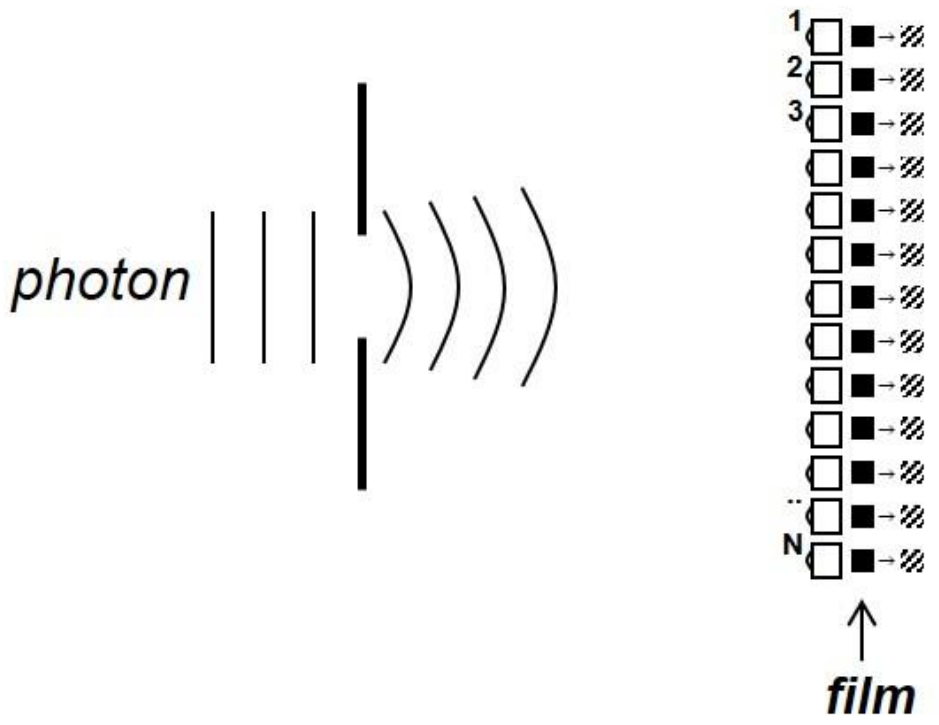

**Figure 6.** Measurement of a photon with a film, which can be modelled as many small mass-displacing detectors.

### 8.2. Reduction condition

In the following, we label the bundle area corresponding to the location of a detector $i$ as $Di$, and denote the local bundles for the cases of "detection" and "no detection" by the detector as $b_i^{Di}$ and $b_{\neg i}^{Di}$. For the intensities of these local bundles, $I_i$ and $I_{\neg i}$, we get $I_{\neg i} = \sum_{j \neq i} I_j$ and $I_i + I_{\neg i} = 1$. The competition action between the two local bundles on $Di$ ($S_{comp\ i,\neg i}^{Di}(\bar{t})$) is abbreviated as $S_{comp}^{Di}(\bar{t})$.

Using norm conservation ($I_i + I_{\neg i} = 1$) and $dI_i + dI_{\neg i} = 0$, the term in the bracket of the conditional equation for collapse (64) can be transformed to $(I_\nu^A dI_\kappa^A - I_\kappa^A dI_\nu^A) = dI_\kappa^A$. Since the inner sum of this equation consists of only one term, it can be transformed based on this result to:

$$dI_i \ = \ \left( \sum_i \frac{S_{comp}^{Di}(\bar{t})}{\hbar} \right) dI_i \, . \tag{69}$$

In the same way as the transformed conditional equation for two-state superpositions in (36), this equation allows for solutions that are not norm-conserving, but are not solutions of the original conditional equation in (64). From (69), it follows that a collapse of the superposition is only possible when the following reduction condition is satisfied:

$$\sum_i S_{comp}^{Di}(\bar{t}_C) = \hbar \, , \tag{70}$$

which is a generalisation of our former result (37) for two-state superpositions.

### 8.3. Localisation of particles during measurement

At the critical spacetime expansion $\bar{t}_C$ (which follows from (70)), the conditional equation in (69) allows for arbitrary solutions for $d\vec{I}_C$. Due to this freedom in the choice of $d\vec{I}_C$, we can choose $d\vec{I}_C$ for each decay-trigger rate $v_{\downarrow trig\ \kappa}^A$ of a local bundle $b_\kappa^A$ individually, in such a way that $d\vec{I}_C$ leads to the decay of that bundle. The final states after collapse resulting from this can be realised by considering the case where only the competition action between two local bundles $b_i^{Di}$ and $b_{\neg i}^{Di}$ on one bundle area $Di$ is non-vanishing. Then, the bundles $b_i^{Di}$ and $b_{\neg i}^{Di}$ must decay as the two paths of a two-state superposition. The decay of bundle $b_i^{Di}$ then leads to a final state $\vec{I}'^{Di}_{\neg i}$, for which the intensity of $b_i^{Di}$ has vanished and the intensity of $b_{\neg i}^{Di}$ has increased, and which is given by:

$$\vec{I}'^{Di}_{\neg i}: \quad I'^{Di}_{\neg i\,k} = \frac{1}{I_{\neg i}^{Di}} \begin{cases} 0\,, & k = i \\ I_k\,, & k \neq i \end{cases} \, . \tag{71}$$

Accordingly, the final state $\vec{I'}_i^{Di}$ in which the intensity of $b_{\neg i}^{Di}$ has decayed and the intensity of $b_i^{Di}$ has increased is given by:

$$\vec{I'}_i^{Di}: \quad I'^{Di}_{i\ k} = \begin{cases} 1\,, & k = i \\ 0\,, & k \neq i \end{cases} . \tag{72}$$

The decay-trigger rates of the bundles $b_i^{Di}$ and $b_{\neg i}^{Di}$ follow from equations (68), (66) and (62):

$$\begin{aligned} v^{Di}_{\downarrow trig\ i}(\bar{t}) \sim \frac{d}{d\bar{t}} S^{Di}_{det\ i}(\bar{t}) = I^{Di}_{\neg i} E^{Di}_{DP}(\bar{t}) \\ v^{Di}_{\downarrow trig\ \neg i}(\bar{t}) \sim \frac{d}{d\bar{t}} S^{Di}_{det\ \neg i}(\bar{t}) = I^{Di}_{i} E^{Di}_{DP}(\bar{t}) \end{aligned} , \tag{73}$$

where $E^{Di}_{DP}$ is the abbreviation for the local Diósi-Penrose energy $E^{Di}_{DP\ i,\neg i}$.

When the spacetime border has reached the critical position $\bar{t}_C$ at which the reduction condition (70) is satisfied, the superposition can decay in an experiment with $N$ detectors in $2N$ different ways. The final states after collapse are the $2N$ local bundles that can be defined for the experiment. The relative transition probabilities for these final states follow from the equations in (73), and yield:

$$\vec{I} \rightarrow \begin{cases} \vec{I'}^{Di}_{i} : & p \sim I^{Di}_{i} E^{Di}_{DP}(\bar{t}_C) \\ \vec{I'}^{Di}_{\neg i} : & p \sim I^{Di}_{\neg i} E^{Di}_{DP}(\bar{t}_C) \end{cases} . \tag{74}$$

Applying our results to the three-detector experiment in Figure 5, we see that the superposition collapses at $\bar{t} = \bar{t}_{C\,1}$, where $\bar{t}_{C\,1}$ is given by $S^{D1}_{comp}(\bar{t}_{C\,1}) + S^{D2}_{comp}(\bar{t}_{C\,1}) + S^{D3}_{comp}(\bar{t}_{C\,1}) = \hbar$ (cf. equation (70)), to one of the six final states $\vec{I'}^{D1}_{1}$, $\vec{I'}^{D1}_{\neg 1}$, $\vec{I'}^{D2}_{2}$, $\vec{I'}^{D2}_{\neg 2}$, $\vec{I}'^{D3}_{3}$ or $\vec{I}'^{D3}_{\neg 3}$ representing the local bundles of the experiment. Figure 7 shows an example of the collapse of the superposition to the final states $\vec{I}'^{D2}_{2}$ or $\vec{I}'^{D2}_{\neg 2}$. For the final state $\vec{I}'^{D2}_{\neg 2}$, in which the system is still in a superposition of paths 1 and 3 (see Figure 7), the system will reduce at a later point in time $\bar{t}_{C\,2}$ to either path 1 or path 3, where $\bar{t}_{C\,2}$ is given by: $S^{D1}_{comp}(\bar{t}_{C\,2}) + S^{D3}_{comp}(\bar{t}_{C\,2}) = \hbar$.

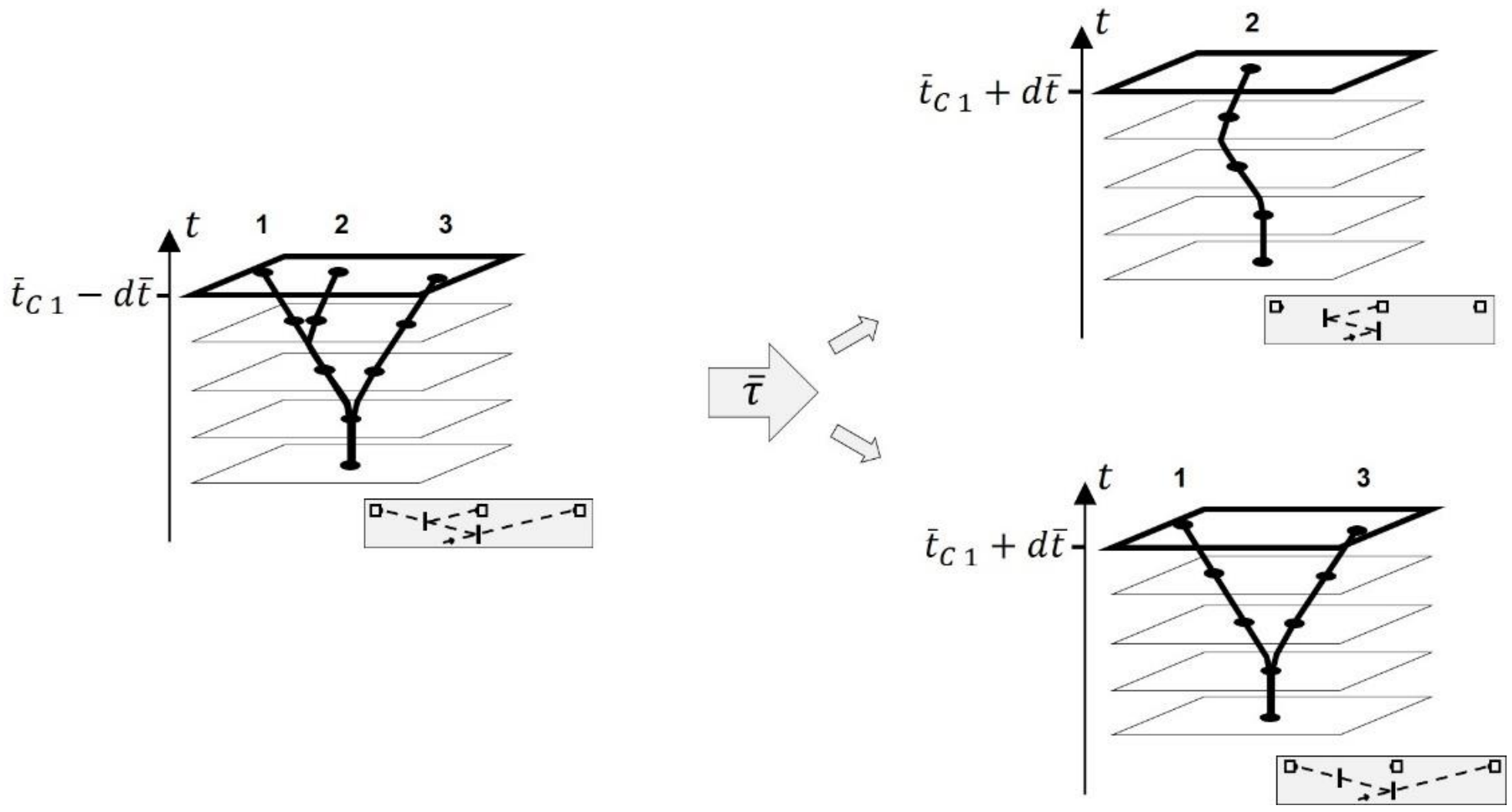

**Figure 7.** Collapse of the evolution of the wavefunction for the three-detector experiment in Figure 5, when the spacetime border reaches the first critical position $\bar{t}_{C\,1}$. Two of the six possible changes are shown (a change in favour of path 2, or a superposition of paths 1 and 3). The insets show the corresponding trajectories of the photon.

At the measurement of a photon by a film, which is modelled in Figure 6 by $N$ small mass-displacing detectors, we get a whole cascade of reductions until the photon is localised at one of the detectors. The first reduction occurs for $\sum_{i=1}^{N} S^{Di}_{comp}(\bar{t}_{C\,1}) = \hbar$. Here, the superposition collapses with a low probability directly to a single state $i$, and with a high probability to a superposition of the other states excluding $i$. In the latter case, the superposition again decays at a later point in time $\bar{t}_{C\,2}$, which is given by $\sum_{k\neq i} S^{Dk}_{comp}(\bar{t}_{C\,2}) = \hbar$, with a low probability to a single state and a high probability to a superposition of the $N-2$ other states. This procedure repeats at subsequent reduction points in time until the superposition is reduced to a single state, which means that the photon is localised at one of the detectors.

### 8.4. Born's rule

From the above discussion of how a photon is localised when it is measured by a film, it follows that the initial superposition can decay in many ways to the same final state. Since the transition probabilities of the single decay steps are proportional to the intensities of the states remaining after the decay, according to (74), it follows that the overall probability $p_i$ for the localisation of a photon at a detector $i$ (i.e. the probability of all paths leading to a localisation of the photon at this detector) is given by Born's rule, i.e. by:

$$p_i = I_i \,. \qquad (75)$$

## 9. Exceptions to Born's rule and faster-than-light signalling

The derivation of Born's rule in Section 8 with help of a property held in common by all experiments performed so far raises the question of whether there are experiments for which our collapse model predicts exceptions from Born's rule. This has the far-reaching consequence of the possibility of faster-than-light signalling, as we will see at the end of this section.

### 9.1. Experiment with three local bundles

Since the derivation of Born's rule in Section 8 is based on the fact that in all of the experiments performed thus far, we never have more than two local bundles at the locations of the detectors, we discuss an experiment in which three local bundles are generated.

In the experiment in Figure 8, a photon is split into three beams, and is measured in beams 1 and 2 by detectors which, when the photon is measured, shift a solid by the distances $\Delta s_1$ and $\Delta s_2$. In beam 0, the photon is not measured, and the solid is not shifted ($\Delta s_0 = 0$). This leads to three local bundles at the location of the solid, which correspond to the displacements $\Delta s_0$, $\Delta s_1$ and $\Delta s_2$ of the solid. To simplify our calculation, we assume that the displacement $\Delta s_1$ is much smaller than $\Delta s_2$ ($\Delta s_1 \ll \Delta s_2$). Since the Diósi-Penrose energy of superposed solids increases typically with the square of the displacement between them ($E_{DP} \sim \Delta s^2$ [44]), the Diósi-Penrose energies between bundles 0 and 2 and between bundles 1 and 2 are then almost identical ($E_{DP\,01} \approx E_{DP\,12}$), and the Diósi-Penrose energy between bundles 0 and 1 can be neglected ($E_{DP\,01} \approx 0$).

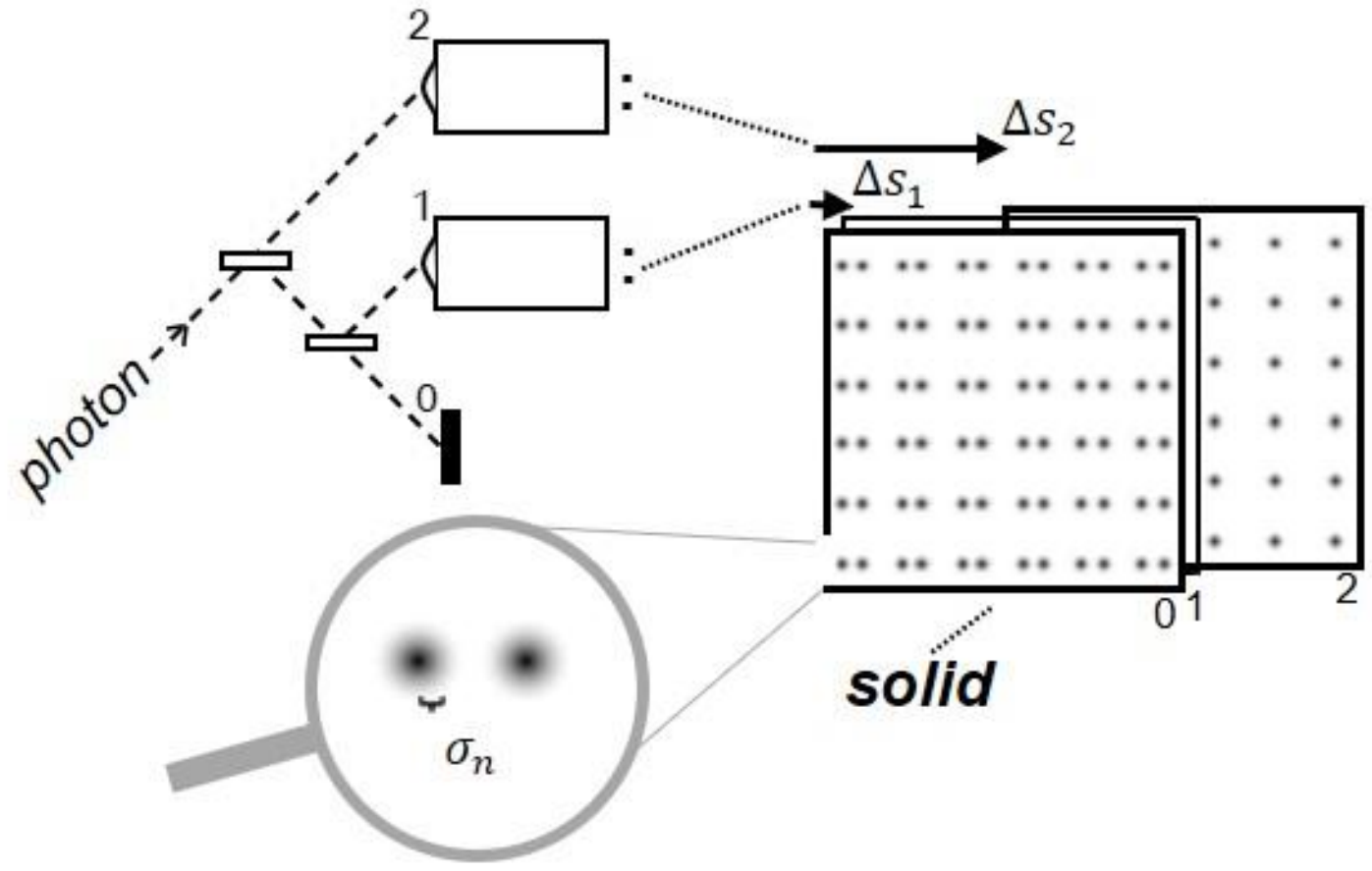

**Figure 8**. Experiment in which three local bundles are generated by transferring a solid into a three-state superposition. For photon detection, detectors 1 and 2 shift the solid by $\Delta s_1$ and $\Delta s_2$, respectively.

When we further assume that the final displacements of the solid are achieved shortly after the photon is split at $t_s = 0$, the Diósi-Penrose energies between the bundles switch after $t_s$ to a constant value, which we abbreviate as $E_{DP}$ ($E_{DP\,02}(t) = E_{DP\,12}(t) = E_{DP}$ for $t > t_s$). The competition actions between the bundles then become $S_{comp\,02}(\bar{t}) = S_{comp\,12}(\bar{t}) = E_{DP}\bar{t}$, and $S_{comp\,01}(\bar{t}) = 0$ for $\bar{t} > t_s$. Inserting this result into the conditional equation for collapse in (64), we get:

$$\begin{pmatrix} dI_0 \\ dI_1 \\ dI_2 \end{pmatrix} = \frac{E_{DP}\,\bar{t}}{\hbar} \begin{pmatrix} I_2 & 0 & -I_0 \\ 0 & I_2 & -I_1 \\ -I_2 & -I_2 & (I_0 + I_1) \end{pmatrix} \cdot \begin{pmatrix} dI_0 \\ dI_1 \\ dI_2 \end{pmatrix} \quad . \tag{76}$$

This equation leads to an expression for the reduction point in time $\bar{t}_C = \hbar / E_{DP}$, with the solution $d\vec{I}_C = (-I_0, -I_1, I_0 + I_1)^T$. Following the discussion in Section 7.3, the three-bundle superposition can collapse at $\bar{t}_C$ to either $\vec{I'}_+ = \vec{I} + \alpha_+ \, d\vec{I}_C = (0,0,1)^T$ or $\vec{I'}_- = \vec{I} - \alpha_- \, d\vec{I}_C = \frac{1}{I_0 + I_1}(I_0, I_1, 0)^T$, representing bundle 2 and the superposition of bundles 0 and 1, respectively. This result intuitively follows from semiclassical gravity, under which superposed states are competing for spacetime geometry. Since bundles 0 and 1 prefer almost the same spacetime geometry, we have a competition only between bundle 2 and the superposition of bundles 0 and 1, which leads to the same reduction point in time ($\bar{t}_C = \hbar / E_{DP}$) as for the two-state superposition (cf. Section 4).

To calculate the reduction probabilities for bundle 2 and the superposition of bundles 0 and 1, we must determine the decay-trigger rates of the three local bundles, which follow from the equations in (67) and (68):

$$\begin{aligned} \nu_{\downarrow trig\,0} &\sim E_{DP}\, I_2 \\ \nu_{\downarrow trig\,1} &\sim E_{DP}\, I_2 \\ \nu_{\downarrow trig\,2} &\sim E_{DP}\, (I_0 + I_1) \end{aligned} \quad . \tag{77}$$

Following the procedure described in Section 7.4 for the calculation of the reduction probabilities, the decay-trigger rates of bundles 0 and 1 both trigger a reduction in favour of bundle 2 (since their intensities decrease when the intensity vector is shifted in $+d\vec{I}_C$-direction), whereas the decay-trigger rate of bundle 2 triggers a reduction in favour of the superposition of bundles 0 and 1. For $p_2 \sim (\nu_{\downarrow trig\,0} + \nu_{\downarrow trig\,1})$ and $p_{(0,1)} \sim \nu_{\downarrow trig\,2}$, we get the following reduction probability for bundle 2:

$$p_2 = \frac{2}{1 + I_2} I_2 \ . \tag{78}$$

For $I_0 = I_1 = I_2 = 1/3$, we get $p_2 = 50\%$, which is higher than predicted by Born's rule ($p_2 = 33{,}3\%$).

Our analysis of the three-bundle experiment in the case where two of the local bundles prefer almost the same spacetime geometry can be summarised as follows. Firstly, the reduction point in time and the final states are identical to those of the related two-state superposition (when merging the two local bundles that prefer the same spacetime geometry into one state); and secondly, the reduction probability of the other bundle (which prefers a different spacetime geometry from the other two) is increased with respect to Born's rule.

### 9.2. Decorrelation of decay-trigger rates

For a deeper understanding of the physical reason why our collapse model leads to reduction probabilities that differ from Born's rule, we analyse the transition from a superposition of two local bundles to a superposition of three local bundles. If we switch off detector 1 in the three-bundle experiment in Figure 8, the solid will evolve into a superposition of only two local bundles. The displacement of the solid is then identical in bundle 0 and 1 ($\Delta s_0 = \Delta s_1 = 0$), and we can combine them into one bundle, which we label as $\overline{01}$ and whose intensity is given by $I_{\overline{01}} = I_0 + I_1$. The decay-trigger rates of bundles $\overline{01}$ and 2 follow from (40) as $\nu_{\downarrow trig\,\overline{01}} \sim E_{DP}\, I_2$ and $\nu_{\downarrow trig\,2} \sim E_{DP}\, I_{\overline{01}}$, which leads to reduction probabilities in agreement with Born's rule ($p_2 = I_2$), as described in the discussion of two-state superpositions in Section 5. Figure 9 visualises the decay-trigger rates $\nu_{\downarrow trig}$ and amplitudes $a_{\downarrow trig}$ for this case (left side) and for the case of three local bundles (right side). The decay-triggers are represented as peaks over the spacetime expansion parameter $\bar{t}$, the amplitudes of which indicate the decay-trigger amplitudes $a_{\downarrow trig}$ and their frequencies the decay-trigger rates $\nu_{\downarrow trig}$. The decay-trigger rates are also visualised by arrows pointing towards the state in favour of which they trigger the decay of the superposition, where the widths of these arrows indicate the amount of the decay-trigger rate. From this figure, it follows that the decay-trigger rate $\nu_{\downarrow trig\ 2}$, which triggers decay in favour of bundle $\overline{01}$ in the two-bundle case and decay in favour of the superposition of bundle 0 and 1 in the three-bundle case, has the same size for both cases. However, in the three-bundle case, the decay-trigger rates triggering a decay in favour of bundle 2 are twice as large as in the two-bundle case, since the decay-trigger rates $\nu_{\downarrow trig\ 0}$ and $\nu_{\downarrow trig\ 1}$ that both trigger a decay in favour of bundle 2 have the same size as $\nu_{\downarrow trig\ \overline{01}}$ (cf. Figure 9), leading to an increased reduction probability of bundle 2.

Figure 9 shows that the decay-trigger amplitudes of bundles 0 and 1 are smaller than for bundle $\overline{01}$, which follows from the fact that the decay-trigger amplitudes are proportional to the intensity of the bundle ($a_{\downarrow trig\ i} \sim I_i$), as discussed in Section 5.3. Hence, the decay-trigger amplitude of bundle $\overline{01}$ is the sum of the decay-trigger amplitudes of bundles 0 and 1 ($a_{\downarrow trig\ \overline{01}} = a_{\downarrow trig\ 0} + a_{\downarrow trig\ 1}$). The decreased decay-trigger amplitudes in the three-bundle case have no impact on the reduction probabilities, since in our collapse model, the smallest intensity fluctuations are sufficient to induce decay of the superposition at the critical spacetime expansion. From the result that the decay-trigger amplitude of bundle $\overline{01}$ is the sum of the decay-trigger amplitudes of bundles 0 and 1, we get an idea of the physical mechanism through which the reduction probability of bundle 2 can change from $p_2 = \frac{2}{1+I_2} I_2$ (for the three-bundle case) to $p_2 = I_2$ (for the two-bundle case) as the displacement between bundles 0 and 1 is continuously decreased ($\Delta s_1 \rightarrow 0$). At this transition, the decay-triggers of bundles 0 and 1 could become correlated and merge into one decay-trigger, whose amplitude is then the sum of the amplitudes of the merged decay-triggers (i.e. $a_{\downarrow trig\ \overline{01}} = a_{\downarrow trig\ 0} + a_{\downarrow trig\ 1}$). The merging of the decay triggers leads to a halving of the decay-trigger rate in favour of bundle 2, which decreases the reduction probability of this bundle to $p_2 = I_2$.This interpretation raises the question of a physical criterion for the decorrelation of the decay-trigger rates of bundles 0 and 1.

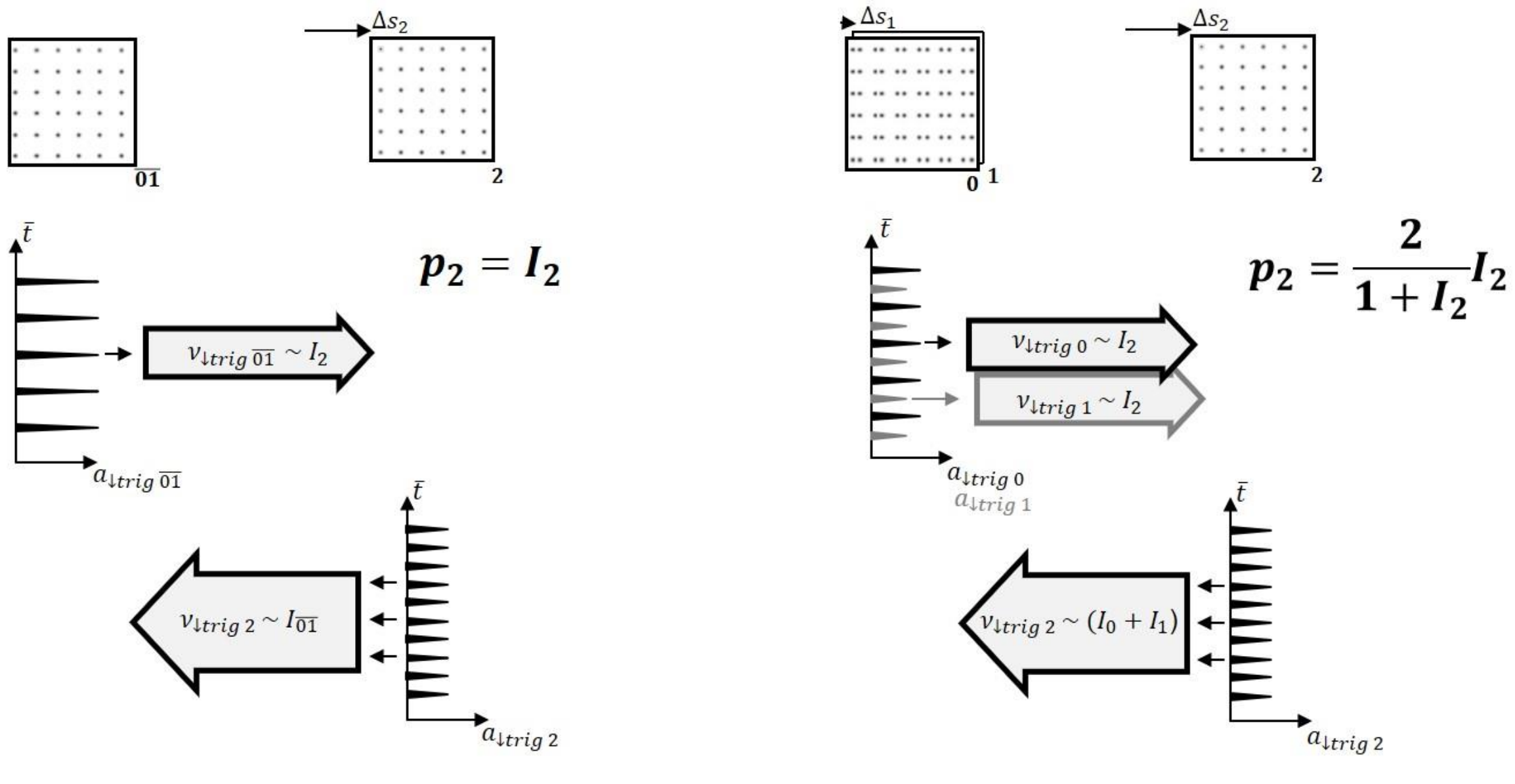

**Figure 9.** Visualisation of the decay-trigger rates $\nu_{\downarrow trig}$ and amplitudes $a_{\downarrow trig}$ of the three-bundle experiment in Figure 8 for the cases where $\Delta s_1 = 0$ (in which states 0 and 1 can be combined into one state $\overline{01}$), and $\Delta s_1 > 0$ (in which the reduction probability of state 2 $p_2$ increases with respect to Born's rule).

### 9.3. Decorrelation criterion

At present, we have no well-founded physical criterion for the decorrelation of the decay-trigger rates, but only a suspicion. For solids in quantum superpositions, we suspect that when the displacement of the solid $\Delta s_1$ between states 0 and 1 is so large at the reduction point in time that the mass distributions of states 0 and 1 are disjoint, this could be sufficient for a decorrelation of their decay-trigger rates. For solids, the mass distributions of their nuclei can be approximately described by Gaussian distributions ($\rho(\boldsymbol{x}) \propto exp(-\boldsymbol{x}^2/(2\sigma_n^2))$, where $\sigma_n$ is the spatial variation of the nuclei $\sigma_n$, as shown in the inset to Figure 8. For a solid at room temperature, the spatial variation $\sigma_n$ is mainly determined by the thermally excited acoustical phonons, and is typically on the order of a tenth of an Ångström [44]. To get disjoint mass distributions for states 0 and 1, the displacement between them must be more than six times larger than the spatial variation in the nuclei $\sigma_n$, and this criterion must be satisfied at the reduction point in time $\bar{t}_C$ ($\Delta s_1(\bar{t}_C) > 6\sigma_n$). For the reduction probability of bundle 2, we get the following result from (42) and (78) for the limiting cases $\Delta s_1(\bar{t}_C) \rightarrow 0$ and $\Delta s_1(\bar{t}_C) > 6\sigma_n$:

$$p_2 = \begin{cases} I_2\,, & \Delta s_1(\bar{t}_C) \rightarrow 0 \\ \frac{2}{1+I_2} I_2\,, & \Delta s_1(\bar{t}_C) > 6\sigma_n \end{cases} . \tag{79}$$

### 9.4. Experimental feasibility and why exceptions to Born's rule have not been observed so far

Although we currently only have a presumption for the decorrelation criterion, it is of interest to discuss on the basis of this criterion the question of whether exceptions from Born's rule can be checked by feasible experiments, and (more importantly) whether exceptions from Born's rule should already have been visible in other experiments, which would put our collapse model in question.

In reference [45], a concrete experiment for transferring a solid into a three-state superposition is proposed, and its feasibility is analysed. From the analysis, it follows that adherence to the decorrelation criterion $\Delta s_1(\bar{t}_C) > 6\sigma_n$ in a concrete experiment requires a new technique for measuring the reduction probabilities. The article in [45] therefore proposed a change of paradigm for measuring reduction probabilities from "reduction by measurement" to "measurement after reduction".

To realise our three-bundle experiment in Figure 8, we must be aware that the equipment used to check which of the detectors has detected the photon is also participating in the quantum superposition, and therefore has an impact on the reduction point in time $\bar{t}_C$. If this equipment shortens the reduction time $\bar{t}_C$ too much, there may be too little time in the time interval from $t_s$ to $\bar{t}_C$ to shift the solid sufficiently far by the distance $\Delta s_1$ to fulfil the decorrelation criterion $\Delta s_1(\bar{t}_C) > 6\sigma_n$. In the experimental proposal in [45], the photon in our three-bundle experiment is measured using avalanche photodiodes, and the displacement of the solid is realised with the help of a piezo, which is steered by the avalanche currents of the photodiodes. The voltages in the avalanche photodiodes are applied with the help of plate capacitors (rather than the usual power supplies), which are charged before measurement with power supplies that are disconnected from the plate capacitors during measurement. This ensures that the power supplies do not participate in the quantum superposition and can shorten the reduction point in time $\bar{t}_C$. Further, the result of the measurement is taken a sufficient time after the superposition has already reduced by connecting voltmeters to the plate capacitors und checking whether their voltages have dropped due to an avalanche current of the connected photodiode (“measurement after reduction”).

A challenge for the quantitative analysis of the experiment is the consideration that not only the solid but also all of the other components of the experiment, including the plate capacitors, the photodiodes, the piezo and the wires connecting these components, participate in the quantum superposition and have an impact on the reduction point in time $\bar{t}_C$. To be able to estimate the impact of these components on $\bar{t}_C$, a formula for the Diósi-Penrose criterion of solids and electric devices in quantum superpositions was developed in [44]. The detailed analysis of the experimental proposal in [45] shows that the decorrelation criterion $\Delta s_1(\bar{t}_C) > 6\sigma_n$ can be ensured when the volume of the solid is quite small ($V \approx 1\, mm^3$). This ensures that reduction does not happen too early ($\bar{t}_C \approx 1\, \mu s$), and the displacement $\Delta s_1$ at the reduction point in time $\bar{t}_C$ is sufficiently large to fulfil the decorrelation criterion $\Delta s_1 > 6\sigma_n$ ($\Delta s_1(\bar{t}_C) \approx 10$Å, $6\sigma_n \approx 0.6$Å, $\Delta s_2(\bar{t}_C) \approx 40$Å).

Since the analysis of the experimental proposal in reference [45] shows that exceptions from Born’s rule can only be observed by new experiments, which must be carefully designed, it becomes obvious why exceptions to Born’s rule have not yet become conspicuous.

### 9.5. Superpositions of many local bundles

Our analysis of the reduction behaviour of three local bundles can easily be extended to more local bundles. For a superposition of $N$ states with equal intensities ($I_i = \frac{1}{N}$), in which one state is distinguished from the others by preferring a different spacetime geometry whereas the others prefer almost identical spacetime geometries, the conditional equation for collapse (64) leads, as shown in Section 9.1 for three local bundles, to two final states. These states are a superposition of those states preferring almost identical spacetime geometries and the other state, which prefers a different spacetime geometry. The decay-trigger rate of the distinguished state is $\nu_{\downarrow trig} {\sim} E_{DP} \frac{N-1}{N}$, whereas the decay-trigger rates of the other states are $\nu_{\downarrow trig} {\sim} E_{DP} \frac{1}{N}$, which all trigger reduction in favour of the distinguished state. When the decay-trigger rates of the other states are decorrelated, the reduction probability of the distinguished state is $50\%$, regardless of how many states $N$ have been considered.

In the event that the exceptions from Born’s rule predicted here are found, it would be of interest to follow up the question of whether nature takes advantage of this effect to reduce a multi-state superposition in favour of a distinguished state with a reduction probability significantly higher than that expected from Born’s rule.

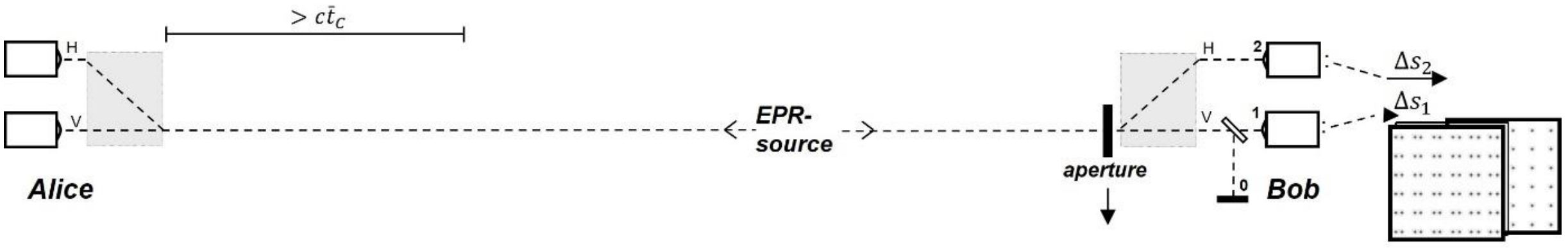

**Figure 10.** EPR experiment for faster-than-light signalling, in which Bob can signal information to Alice by changing the probability of the H-polarisation from $p_H = \frac{1}{2}$ to $p_H = \frac{2}{3}$ by removing the aperture.

## 9.6. Faster-than-light signalling

The prediction of our collapse model in terms of the design of experiments in which the reduction probabilities may differ from Born's rule has a far-reaching consequence, which is the option to construct experiments involving faster-than-light signalling. This prediction of the Temporally Expanding Spacetime approach does not provoke a conflict with relativity, since the following discussion of the signalling experiment refers only to the property of our collapse model in which the evolution of the wavefunction can change abruptly at the critical spacetime expansion, and the evolution of the wavefunction before and after this abrupt change is governed solely by local physical laws.

Figure 10 shows how an EPR experiment can be enhanced for faster-than-light signalling. The EPR source of the experiment generates the Bell state $|\psi > \sim |H > |H > + |V > |V >$. In this experiment, Bob has the option to change the reduction probabilities of the states $|H > |H >$ and $|V > |V >$ from the usual $p_H = p_V = \frac{1}{2}$ to $p_H = \frac{2}{3}$, $p_V = \frac{1}{3}$ with the setup of our three-bundle experiment in Figure 8, when he removes the aperture from the photon beam, as shown in Figure 10. Then, the solid in the setup is shifted by $\Delta s_2$ for detection of the $|H > |H >$-state, and in the other case is split into two states that are shifted with regard to each other by $\Delta s_1$. This increases the reduction probability of the $|H > |H >$ -state from $p_H = \frac{1}{2}$ to $p_H = \frac{2}{3}$ (cf. (78)). To guarantee that Alice can observe this changed reduction probability, it is important to ensure that the superposition is reduced only by Bob's measurement, which can be achieved by setting Alice's arm to at least $c\bar{t}_C$ longer than Bob's, as shown in Figure 10. Bob can then signal information to Alice by manipulating the reduction probabilities of the superposition.

# 10. Discussion

The proposed Temporally Expanding Spacetime approach to wavefunction collapse, based on the hypothesis of a temporally expanding spacetime and semiclassical gravity, can explain the collapse of the wavefunction, can resolve the conflict between quantum nonlocality and relativity, predicts the same lifetimes of superpositions as the gravity-based approaches of Diósi and Penrose, and is consistent with the facts that all experiments performed thus far behave in agreement with Born's rule and that there is no evidence that energy might not be conserved at collapse. Furthermore, this approach predicts new behaviour in the form of exceptions to Born's rule and faster-than-light signalling, which could enable its verification.

## 10.1. Temporally expanding spacetime

The hypothesis that spacetime might have a temporal border that is dynamically moving towards the future was previously proposed by Muller in the context of cosmology [57]. Muller assumes that the temporal expansion of the universe started with its spatial expansion at the Big-Bang singularity, and that the propagation of the temporal border is aligned with the rest frame of background radiation. Muller's hypothesis is motivated by the need to address the missing concept of the "now" in physics, which he defines by the position of the temporal border. In the Temporally Expanding Spacetime approach, we can even define a temporal depth for this "now" based on the spacetime regions in which the state vector is still in superposition, and in which the state vector and accordingly the mass distribution can abruptly change due to collapse events.

Muller also emphasises the implications of his hypothesis for the discussion of the "arrow of time" in physics, which make this arrow explicit, since the past belongs to the spacetime region before the temporal border and future events will happen in spacetime regions that the temporal border has not yet reached. It is of interest to mention that the Temporally Expanding Spacetime hypothesis provides an explanation for why advanced potentials (such as in electrodynamics) cannot play a role and must be left out of the solutions.

## 10.2. Faster-than-light causality

The most important prediction of the Temporally Expanding Spacetime approach is that the evolution of the wavefunction can change abruptly across entire spacetime regions, leading to faster-than-light causality, which can explain the quantum correlations in EPR experiments. This explanation of collapse agrees with the conclusion of the Free Will theorem of Conway and Kochen, which showed that in special EPR experiments with free choice of measurements (in the sense that the choices are not functions of the past), information about the choices must travel infinitely fast between the measurement partners [22-24]. This result led to a scientific debate over whether it is possible in principle to define relativistic collapse models [25-28]. Our analysis of the Temporally Expanding Spacetime approach to wavefunction collapse has shown that faster-than-light causality must not lead to a conflict with relativity when the dynamics of physics is defined in a new way.

However, the Temporally Expanding Spacetime has far exceeded its objective of resolving the conflict between quantum nonlocality and relativity by also predicting faster-than-light signalling. If this prediction of the Temporally Expanding Spacetime approach is confirmed, it would have profound implications for other areas of physics.

### 10.3. Comparison with the models of Diósi and Penrose

The fact that the Temporally Expanding Spacetime approach leads to the Diósi-Penrose criterion in the same way as the approaches of Diósi and Penrose affirms its importance for gravity-based approaches to wavefunction collapse.

An important feature of our approach is that the reduction point in time is not expected to vary statistically. Although statistical variation in the reduction point in time is not explicitly discussed for the models of Diósi [6,7] and Penrose [9,10], their approaches suggest such a statistical variation.
A further important characteristic of the Temporally Expanding Spacetime approach is that energy is conserved at collapse. In Diósi's model, collapse events lead to a permanent increase in energy [7,8], which could lead to spontaneous photon emissions [58], although these have not been observed so far [59]. To avoid an unrealistic high increase in energy, Diósi had to introduce a modification of the mass-density operator in form of a smearing of the mass density [7], which leads to longer lifetimes of superposed solids when the displacement between them is small [44].

### 10.4. Experimental verification

If the experimental problems related to measuring the collapse of the wavefunction can be solved in the future, the question of whether gravity is the physical origin of collapse can be answered by reference to the Diósi-Penrose criterion. The differences between the Temporally Expanding Spacetime approach and the models of Diósi and Penrose described in the previous paragraph might then help to determine which of these gravity-based approaches describes nature best.

However, experimental checks of the Temporally Expanding Spacetime approach must not be delayed. The prediction of exceptions to Born's rule leads to an experimental agenda that can be carried out in the short term. The experimental proposal for transferring a solid into a three-state superposition given in reference [*45*][4] (cf. Section 9.4) can be realised with commercially available equipment, and can be carried out at room temperature. Although we currently have only a suspicion of the criterion for the decorrelation of the decay-trigger rates leading to deviations from Born's rule (cf. Section 9.3), the Temporally Expanding Spacetime approach gives concrete guidance on where and how to search for exceptions to Born's rule.

If deviations from Born's rule can be found, an agenda for checking the further predictions of the Temporal Expanding Spacetime approach should be pursued. The feasibility of the EPR experiment for faster-than-light signalling (Section 9.6) is also demonstrated in reference [45]. For a reduction time of $0.84\,\mu s$, Alice's arm must be chosen to be at least $250\,m$ longer than Bob's, in order to ensure that the superposition is reduced by his measurement.

[4] Reference [45] uses a slightly different notation than here, since the paper is based on a previous version of our model [43,42], but which leads to the same formulas.

Deviations from Born's rule would also allow us to construct experiments for indirectly measuring the lifetimes of superpositions [45]. If for the three-bundle experiment in Figure 8, we introduce a time-delay for the displacement $\Delta s_1$, so that at time $t_s$ the solid first evolves into a superposition of two bundles, consisting of bundle $\overline{01}$ and 2, and only after the time-delay, at time $t_{s\,\overline{01}}$, into a superposition of three bundles, we observe an increased reduction probability for bundle 2 only if the superposition of bundle $\overline{01}$ and 2 has not already reduced before $t_{s\,\overline{01}}$ (i.e. $p_2 = I_2$ for $\bar{t}_C < t_{s\,\overline{01}}$, $p_2 = \frac{2}{1+I_2} I_2$ for $\bar{t}_C > t_{s\,\overline{01}}$). Such measurements would allow us to check the Diósi-Penrose criterion and check whether the reduction point in time varies statistically or not.

There is even the possibility of measuring the relative velocity of the spacetime border with respect to the reference frame of the Earth. For a sufficiently large distance between Alice and Bob in the signalling experiment in Figure 10, there are cases in which the spacetime border passes Alice's measurement before Bob's. In this case, Bob can no longer manipulate the reduction probabilities of the superposition and signal information towards Alice. For a relative velocity of the Earth with respect to the reference frame of background radiation of $v = 369\ km/s$ [60] and a reduction time of $\bar{t}_C = 0.1\,\mu s$ [44], the distance between Alice and Bob must be about $50\ km$ ($d \approx \frac{2c^2}{v} \bar{t}_C$) to allow us to observe this effect.